\documentclass[a4paper,fleqn,usenatbib]{mnras}
\usepackage{ae,aecompl}
\usepackage{graphicx}	% Including figure files
\usepackage{amsmath}	% Advanced maths commands
\usepackage{amssymb}	% Extra maths symbols
\usepackage{subfigure}

\title[\texttt{OASIS} - A New Planet Simulator]{Modelling the 3D Climate of Venus with \texttt{OASIS}}

\author[Mendon\c ca $\&$ Buchhave]{
Jo\~ao M. Mendon\c ca$^{1}$\thanks{E-mail: joao.mendonca@space.dtu.dk}\thanks{Homepage: software-oasis.com}, Lars A. Buchhave$^{1}$
\\
$^{1}$National Space Institute, Technical University of Denmark, Elektrovej, 2800, Kgs. Lyngby, Denmark.\\
}

\date{Accepted XXX. Received YYY; in original form ZZZ}

\pubyear{2019}

\begin{document}
\label{firstpage}
\pagerange{\pageref{firstpage}--\pageref{lastpage}}
\maketitle

% Abstract of the paper
\begin{abstract}
Flexible 3D models to explore the vast diversity of terrestrial planets and interpret observational data are still in their early stages. In this work, we present \texttt{OASIS}: a novel and flexible 3D virtual planet laboratory. With \texttt{OASIS} we envision a platform that couples self-consistently seven individual modules representing the main physical and chemical processes that shape planetary environments. Additionally, \texttt{OASIS} is capable of producing simulated spectra from different instruments and observational techniques. In this work we focus on the benchmark test of coupling four of the physical modules: fluid dynamics, radiation, turbulence and surface/soil. To test the \texttt{OASIS} platform, we produced 3D simulations of the Venus climate and its atmospheric circulation and study how the modelled atmosphere changes with various cloud covers, atmospheric heat capacity, and surface friction. 3D simulations of Venus are challenging because they require long integration times with a computationally expensive radiative transfer code. By comparing \texttt{OASIS} results with observational data, we verify that the new model is able to successfully simulate Venus. With simulated spectra produced directly from the 3D simulations, we explore the capabilities of future missions, like LUVOIR, to observe Venus analogues located at a distance of 10 pc. With \texttt{OASIS}, we have taken the first steps to build a sophisticated and very flexible platform capable of studying the environment of terrestrial planets, which will be an essential tool to characterize observed terrestrial planets and plan future observations.
\end{abstract}

\begin{keywords}
Hydrodynamics -- methods: numerical -- Planets and satellites: atmospheres 

\end{keywords}

%%%%%%%%%%%%%%%%%%%%%%%%%%%%%%%%%%%%%%%%%%%%%%%%%%

%%%%%%%%%%%%%%%%% BODY OF PAPER %%%%%%%%%%%%%%%%%%

\section{Introduction}
\subsection{Background}
The research field of planetary sciences has significantly broadened its reach and scope in the last two decades with the discovery of a multitude of planets orbiting other stars. There are more than 4,000 exoplanets known but the variety of environments that they may harbour is still unknown. Most importantly, it propels the fundamental scientific and philosophical quest of searching for the first detection of life beyond our own planet. Recently, results from the Kepler Space Telescope suggest that there are at least two Earth-size planets orbiting every M-dwarf star (\citealt{2015Dressing}), which are the most abundant stars in our galaxy. M stars are colder and less massive than our Sun, which allows planets lying closer to their parent star to still retain atmospheres, and in some cases, to have the right conditions to host liquid water at their surfaces (\citealt{2013Forget}). Being closer to their stars also facilitates probing their atmospheres with current observational methods. Recently, terrestrial planets orbiting M stars with potential habitable conditions have been detected, such as, the TRAPPIST-1 planet system (\citealt{2016Gillon, 2017Gillon}; \citealt{2017Luger}), Proxima Centauri b \citep{2016Anglada}, LHS 1140b \citep{2017Dittmann} and Ross 128b \citep{2018Bonfils}. Space missions such as the James Webb Space Telescope (JWST), which will be launched in 2021, will be able to detect water features in the atmosphere of, for example, Earth-size planets around M-dwarf stars (\citealt{2017Morley}; \citealt{2018Batalha}; \citealt{2019Lustig-Yaeger}; \citealt{2019Wunderlich}). However, a theoretical framework is needed to support these observations. Otherwise, the observational data may not be correctly interpreted, causing the community to miss possible habitable planets or erroneously identify barren planets as habitable. Furthermore, without a theoretical guide to select planet candidates, valuable observational resources may be wasted on planets, which theoretically are unlikely to harbour life.

Meaningful characterizations of planetary environments will require a large effort from theory and observations. Our chances of finding habitable planets and tracers of life may be increased if we effectively optimize the telescope observing time to target the most likely habitable planets and if we are able to accurately interpret the ensuing observational data. The best targets are planets with long term stability in their environment allowing liquid water at the surface (usually defined as habitable conditions, e.g., \citealt{2013Forget}). This requires robust knowledge of the time evolution of planetary climates. To understand how planet environments evolve we need to characterize the main mechanisms driving the climate and to explore the interplay of the chemical and physical processes using numerical models. 

The characterization of exoplanets environments will be accomplished via remote sensing, where the theoretical platforms will have an important role interpreting the planet spectra and variations of the observed signals. Planetary environments can be very complex and a deep understanding of climate problems often requires a hierarchy of theoretical models. This is necessary due to the difficulty in interpreting results that are mostly associated with the intrinsic non-linear nature of the problems. Such a hierarchical strategy allow us to gather important information on the assumptions, concepts and limitations of the models developed. At the end of this hierarchy of models we have the 3D Global Circulation Models (GCMs). In this work we focus on this last category. GCMs are powerful tools that include self-consistent representations of the main physical and chemical processes that drive the planet's environment (e.g., \citealt{2008Donner}). GCMs allow researchers to study the dynamical transport of heat, chemistry and clouds across the 3D atmosphere. These advantages allow us to more accurately obtain the global temperature of planetary atmospheres. Several works have shown that 3D effects, such as, a variation of the planet's albedo due to the sensitivity of the atmospheric circulation and the sub-stellar cloud deck to the planet's rotation rate (\citealt{2016Kopparapu}) or the cold trapping of volatiles in the permanent night side of tidally locked terrestrial planets (\citealt{2013Leconte}) are important to understand the planet's environment. The heat transport in the atmosphere of tidally-locked planets (these planets have permanent day- night-sides) can have a very strong impact on the planet's environment, which in the case of inefficient heat transport can lead to atmospheric collapse (e.g., \citealt{2015Wordsworth}). The terrestrial planets around low mass stars in the habitable zone are potentially in a tidally locked state and the exploration of the heat transport in these scenarios cannot be done self-consistently with 1D models. Due to these features, 3D GCMs are one the best and most reliable tools to study the environment in these planets. The observed data contains inherently 3D information on the planetary atmospheres that can be misinterpreted by simpler 1D models, which are unable to robustly represent the impact of the atmospheric circulation. Recently, \cite{2019Caldas} showed that day-night thermal and compositional contrast in tidally locked planets can produce a gradient in opacities which has an important impact on the planet spectrum. Data from thermal emission and reflectance spectra will also be strongly modulated by 3D effects caused by longitudinal differences in, for example, temperature, cloud cover and surface/oceans. There is a long list of works that have used GCMs to study terrestrial planets with different levels of complexity (e.g., \citealt{2019Shields}). The terrestrial planets studied with GCMs range from our own Earth, to planets in our Solar System and beyond. The diversity of studies used to explore planetary environments using 3D climate models include, for example, faint young Sun paradox on Earth and Mars (e.g., \citealt{2013Charnay} and \citealt{2013Wolf}, \citealt{2017Wordsworth}), planetary snowball episodes (e.g., \citealt{2010Abbot}), Mars atmosphere (e.g., \citealt{1999Forget}), Venus atmosphere (e.g., \citealt{2010Lebonnois} and \citealt{2016Mendoncaa}), Titan atmosphere (e.g. \citealt{2015Charnay}), Pluto atmosphere (e.g., \citealt{2017Forget}) or evaluate the climate of terrestrial exoplanets (e.g., \citealt{2011Wordsworth}, \citealt{2013Leconte}, \citealt{2016Shields}, \cite{2017Turbet}, \cite{2018HaqqMisra}, \citealt{2019Komacek} and \citealt{2019DelGenio}). GCMs to study terrestrial climates have also included the effect of atmosphere-ocean coupling, and have shown that ocean dynamics play an important role in the characterization of planet climates in the middle range and outer edge of the habitable zone (e.g., \citealt{2019Yang}). The disadvantage of GCMs against other simplified models such as the 1D radiative-convective models, is the heavy computations required for the 3D calculations. However, progress has been made to boost the performance of 3D calculations such as the implementation of the 3D code to run on Graphic Processing Units (GPUs) that benefits from the GPUs massively parallel architecture (\citealt{2016Mendoncab}).

Upcoming missions will offer exciting opportunities to implement new techniques to characterize terrestrial planets and search for habitable atmospheres. Combining robust theoretical platforms and observations will be essential to formulate the best methods for the first rigorous characterization of terrestrial planets.

\subsection{Motivation}
\label{subsec:Motiv}
With the aim of exploring the large diversity of possible terrestrial climates we are currently developing the platform \texttt{OASIS}\footnote{Recent news and updates on the \texttt{OASIS} platform can be found in \href{www.software-oasis.com}{www.software-oasis.com}.}. \texttt{OASIS} is a 3D virtual lab that will include the representation of the main physical and chemical processes that shape planetary climate and their evolution. The new platform has been written completely from the ground up to avoid approximations that could compromise the flexibility of the model to explore a large diversity of planetary conditions. The model has also been developed to run on GPUs (Graphics Processing Units) in order to be very efficient exploring a large parameter space of planetary characteristics, using complex and computationally expensive physical/chemical routines, and be able to do long time integrations. We are also developing tools that will use the 3D \texttt{OASIS} output to simulate the observed spectra for different observational methods. Our goal is for this platform to play a key role in helping the exoplanet community identify important targets for follow-up observations.

In this work our goal is to present the first results of our new platform. We focus mainly on the coupling between the dynamical core (\texttt{THOR}), atmospheric turbulence (\texttt{LOKI}), surface/soil thermodynamics (\texttt{ATLANTIS}) and the radiative transfer (\texttt{CYCLOPS}). The numerical methods in these four modules are benchmarked in this work by computing the 3D Venus-like environment. As we explain later, the aim of \texttt{OASIS} is to include more self-consistent physical and chemical schemes, such as cloud formation and atmospheric chemistry. However, the representation of clouds and chemistry in this work is very simplified (the model assumes that the cloud cover and atmospheric composition are constant as a function of time). Our goal is not to tune the model to simulate accurately the Venus atmosphere, but to select the main Venus bulk parameters, and test whether our numerical model can reproduce the main properties of the Venus atmospheric circulation and temperature structure without further input. By exploring the atmosphere of Venus we test \texttt{OASIS} with the most computationally challenging terrestrial planet to simulate in our Solar System. To be able to simulate Venus, the numerical models must be able to simulate the global atmosphere for thousands of Earth days until the model converges to a steady state (e.g., \citealt{2010Lebonnois} and \citealt{2016Mendoncaa}). The long integration is related to the large thermal inertia of the atmosphere that is weakly forced by stellar radiation (just about 2.5$\%$ of the incoming stellar radiation reaches the surface (e.g., \citealt{1980Tomasko}, \citealt{2015Mendonca}). The radiation processes play an important role driving the dynamics and climate. To accurately represent the radiative processes in Venus it requires a computationally expensive radiation transfer scheme due to the optically thick massive atmosphere that is covered with highly reflective clouds (e.g., \citealt{2009Eymet}). The radiation scheme also needs to be able to represent well the spectral windows that allow a more efficient energy exchange between the deep atmosphere and the upper layers above the clouds. 

The challenging problems pointed out above set Venus as a good benchmark test to the coupling between the dynamical core (\texttt{THOR}), atmospheric turbulence (\texttt{LOKI}), surface/soil thermodynamics (\texttt{ATLANTIS}) and the radiative transfer (\texttt{CYCLOPS}). We also want to explore the impact of some important model parameters on the simulated atmospheres (cloud cover, atmospheric heat capacity and surface friction). Having a robust model for Venus, allows us to have confidence when characterizing Venus-like planets. Transit detections favour planets orbiting closer to the stars (e.g., \citealt{2008Kane}, \citealt{2014Kane}), which may be an indication that we will find more planets with atmospheric conditions similar to Venus than similar to Earth or Mars. Also, exploring Venus-like planets allow us to learn more about the inner edge of the habitable zone, and improve our understanding on the climate evolution of Earth and Venus.

Our long-term goals are to transform \texttt{OASIS} into a platform capable of modelling the atmospheric composition and evolution of terrestrial planets, and understand the key physical processes behind the diversity of planetary environments.

\subsection{Venus}
\label{sec:venus}
Venus is the most Earth-like planet in the Solar System in terms of mass and size. However, the current state of both planetary environments is very distinct. If the two planets started with similar planet conditions, different processes shaped distinct climate evolutions: stronger incoming stellar radiation in Venus (closer to the Sun), no magnetic field in Venus which makes atmospheric escape more efficient and lack of a geochemical cycle in Venus that could remove volatiles such as CO$_2$ into the planet's interior. The state of the early Venus climate is still controversial due to lack of observational constraints and robust theory to support different scenarios. Some studies suggest that early Venus had a temperate climate that allowed liquid water to be stable at its surface, and that the planet lost all its oceans to the atmosphere during a net positive feedback between the surface temperature and the atmospheric opacity that enhances the strength of the greenhouse effect (\citealt{1969Ingersoll}). Another possibility is that early Venus had a massive steam-filled atmosphere where the water could never condense at its surface (\citealt{2013Lebrun}).

The current Venus has a massive atmosphere composed mostly of CO$_2$ that creates a hostile environment at its surface: temperatures reaching roughly 735 K and pressure 92 bars (\citealt{2018Taylor}). 96.5$\%$ of the dense atmosphere is CO$_2$ and 3.5$\%$ N$_2$ including minor tracers such as 30 ppm of H$_2$O and 150 ppm of SO$_2$. The large quantity of CO$_2$ is responsible for the large temperatures obtained at the surface (``greenhouse'' phenomenon, \citealt{1961Sagan}). Other minor constituents in the atmosphere of Venus have been observed such as, water, sulphur dioxide and carbon monoxide (\citealt{2018Taylor}). The planet is completely covered by clouds that have a big impact on the energy budget of the atmosphere. Venus' clouds are mostly a mixture of sulphuric acid and water droplets, extending from an altitude of $\sim$45 to \mbox{65 km}, with layers of sub-micron particles below and above the main cloud deck (\citealt{1980Knollenberg}). The clouds of Venus reflect most of the incoming stellar radiation back to space (the bond albedo is $\sim$0.75), and are very opaque in the UV, visible and most of the infrared wavelength range (\citealt{2016Read}). However, there are some visible and near-IR spectral windows that have been used to observe the Venus lower atmosphere (e.g., \citealt{1984Allen}). There is a strong absorber in the cloud region in the UV range. This absorber is easily detected in the Venus UV images due to its effective absorption and inhomogeneous spatial distribution across the atmosphere (e.g., creates a horizontal ``V'' shape structure across the atmosphere). The composition of this UV absorber is still not confirmed and some of the proposed identities of the unknown UV absorber are for example: sulphur aerosols, FeCl$_3$ (\citealt{2017Krasnopolsky}),  sulphur oxide isomer (OSSO, \citealt{2016Frandsen}) or even organic components (\citealt{2018Limaye}). The UV absorber has also an important role in the deposition of stellar energy across the atmosphere.

Venus has almost no seasons (the spin inclination axis is $\sim$2.64$^\circ$) and an orbital eccentricity around 0.007. The solid planet rotates slowly in a retrograde direction, and takes roughly 240 days to complete a full revolution (\citealt{1964Goldstein} and \citealt{1964Carpenter}). Despite the solid planet rotating slowly the atmosphere is rotating faster in a phenomenon called super-rotation. The cloud deck is estimated to take on average 4-5 Earth days to rotate around the planet (50-60x faster than the solid planet). The winds reach speeds of 100 m$/$s in the cloud region, and the mechanisms for their formation are described in \cite{2010Lebonnois} and \cite{2016Mendoncaa} as a combination of zonal mean circulation, thermal tides and transient waves, where the semi-diurnal tide plays a crucial role transporting angular momentum mainly from the upper atmosphere towards the cloud region. The super-rotation index ($S$) or Read number quantifies the magnitude of the super-rotation phenomenon in a planetary atmosphere, which is defined by the excess of total angular momentum of the atmosphere ($M_t$) compared to an atmosphere co-rotating with the solid planet($M_0$):
\begin{equation}
S = \frac{M_t}{M_0}-1.
\label{eq:s}
\end{equation}
Eq. \ref{eq:s} was first defined in \cite{1986Read}. \cite{2016Mendoncaa}, estimate from observational wind measurements (\citealt{1985Kerzhanovich}) a Read number of 8.7$^{+5.2}_{-4.6}$ for Venus. Further details on how to compute the Read number can be found in \cite{1986Read} or \cite{2018Mendoncab}.

\subsection{Structure of the present study}
In this paper we present the structure of what will be the \texttt{OASIS} platform and the first tests on the different modules working together. In Section \ref{sec:OASIS}, we present the different physics modules at work in \texttt{OASIS}. In Section \ref{sec:ref_sim}, the results of the reference simulations with different spatial resolutions are presented and discussed. The results of the reference simulations are compared against the observations in Section \ref{sec:obs}. In Section \ref{sec:sen_tests} we perform sensitivity tests to the reference simulation where we explore a different cloud cover (\ref{sec:clouds}), atmospheric heat capacity (\ref{sec:cp}) and surface friction (\ref{sec:fri}). In Section \ref{sec:venus_exoplanet}, we move our simulated Venus planet to orbit a Sun-like star at 10 parsec away, and explore the results using an instrument with capabilities similar to the LUVOIR mission concept. Finally, in Section \ref{sec:conc}, we discuss our model results and future perspectives.  

\begin{figure*}
\begin{centering}
\includegraphics[trim={0cm 0cm 0cm 0cm},clip,width=1.5\columnwidth]{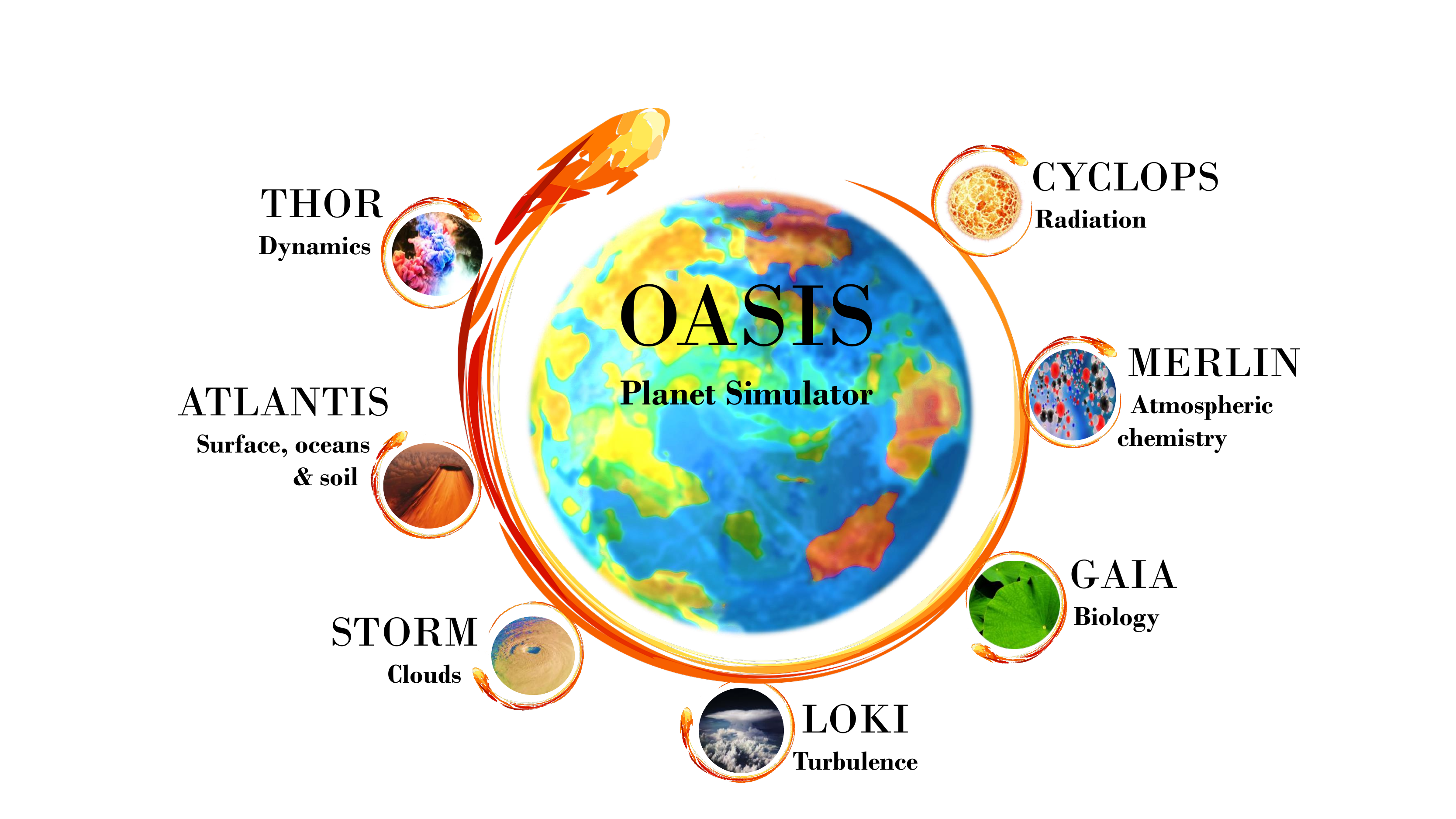}
\caption{The architecture of the \texttt{OASIS} platform. The different modules around \texttt{OASIS} represent the different physical and chemical processes of the 3D platform: \texttt{THOR} represents the dynamical fluid flow in the atmosphere; \texttt{CYCLOPS} the interaction of the radiation with the atmosphere and surface; \texttt{LOKI} the turbulence in the atmosphere; \texttt{STORM} the cloud physics; \texttt{ATLANTIS} the mechanical interaction between the atmosphere and the surface, and the thermodynamics of the planet soil; \texttt{MERLIN} represents the atmospheric and surface chemistry; \texttt{GAIA} is planned to be a module that will represent the effects of biology on the surface planet and on its atmosphere.}
\label{fig:oasis}
\end{centering}
\end{figure*}

\section{\texttt{OASIS} - A dedicated Planetary model}
\label{sec:OASIS}
\texttt{OASIS} is a new theoretical platform developed to expand our knowledge of planetary environments. The novel planetary virtual-lab is aimed to couple self-consistently seven modules (Fig. \ref{fig:oasis}) that represent physical and chemical processes in the planets, and it can be used as a fast 1D model (see the \texttt{MIRAGE} module below) or full 3D. This platform is part of a long term project that gradually will include more physically-based and complex parametrizations, and also model options to control the complexity of the model. 

In the core of \texttt{OASIS} is \texttt{THOR} (\citealt{2016Mendoncab}) - a module that includes a state-of-the-art dynamical solver, which represents the atmospheric fluid flow. Coupled self-consistently with \texttt{THOR} we have currently three physical modules and three that are currently being developed:
\vspace{0.08cm}
\newline
\underline{Completed}
\begin{description}
\item[\texttt{CYCLOPS}] represents the absorption and scattering in the atmosphere and surface.
\item[\texttt{LOKI}] parametrizes the small-scale physics that requires very fine space and time resolutions to be fully represented (e.g., turbulence).
\item[\texttt{ATLANTIS}] represents the physics at the planet surface, soil, and the impact of the surface boundary in the atmosphere.
\end{description}
\underline{Under development}
\begin{description}
\item[\texttt{STORM}] incorporates our physical scheme for cloud formation and transport.
\item[\texttt{MERLIN}] solves the chemical equations in the atmosphere and surface.
\item[\texttt{GAIA}] is a module that is currently being developed and includes the manifestation of the presence of biology in the planet atmosphere and surface.
\end{description}

\texttt{OASIS} is being gradually improved in terms of code performance, usability and complexity in the physical/chemical modules. \texttt{OASIS} has a modular configuration that allow us to select the parts of the code that we want to include in the simulations. In the next subsection we describe the current content of each completed modules and assumptions about gas composition and clouds. Note that the module names in \texttt{OASIS} do not correspond to acronyms. The main purpose of the module names is to allow the model user to easily remember the existing different model options and to highlight that each module can work as a standalone code or integrated to work together with other modules.

\begin{table}
\begin{center}
\begin{tabular}{ | l | l | l | l |}
\hline
 Parameters & Venus & Earth & Units  \\ \hline \hline
 Planet distance & 0.7 & 1 & AU \\ \hline
 Mean Radius & 6052 & 6371 & km \\ \hline
 Gravity (equator) & 8.87 & 9.78 & m/s$^2$ \\ \hline
 Gas constant & 188 & 287 & J K$^{-1}$kg$^{-1}$ \\ \hline
 Specific heat & 900 & 1005 & J K$^{-1}$kg$^{-1}$ \\ \hline
 Rotation rate & -2.99$\times$10$^{-7}$ & $7.29\times {10}^{-5}$ & s$^{-1}$ \\ \hline
 \end{tabular}
\end{center}
\caption{Model parameters used in the baseline simulation of Venus.}
\label{tab:model}
\end{table}

\subsection{Completed modules}
\subsubsection{\texttt{THOR} - Dynamical Core}
\label{subsec:thor}
\texttt{THOR} is a state-of-the-art atmospheric dynamical core that solves the three dimensional non hydrostatic Euler equations on a modified icosahedral grid (\citealt{2016Mendoncab} and \citealt{2019Deitrick}). The core architecture of \texttt{OASIS} is developed around this novel model. \texttt{THOR} is part of the Exoplanet Simulation Platform\footnote{ESP has a variety of atmospheric tools open-source, such as: \texttt{VULCAN}: chemical kinetics for atmospheres (\citealt{2017Tsai}); \texttt{HELIOS}: radiative transfer code (\citealt{2017Malik}, \citealt{2019Malik}); \texttt{HELIOS-R}: nested sampling atmospheric retrieval code (\citealt{2017Lavie}); \texttt{HELIOS-T}: atmospheric retrieval code for exoplanet transmission spectra (\citealt{2018Fisher});} (ESP) tools, and it can be downloaded at https://github.com/exoclime. The main advantages of using \texttt{THOR} against other recent planetary models is that:
\begin{itemize}
\item The atmospheric fluid flow is completely represented, and no approximations are used that could compromise the physics. In this work \texttt{THOR} solves the 3D nonhydrostatic compressible Euler equations on a rotating sphere (\citealt{2016Mendoncab}). \texttt{THOR} uses the Horizontally Explicit Vertically Implicit (HEVI - \citealt{2002Satoh}) method to allow an increased integration time-step which would otherwise be constrained by fast sound waves.
\item The equations are solved on an icosahedral grid that avoids the problem associated with the convergence of the meridians at the poles in latitude-longitude grids.
\item The interface is user-friendly and can be easily adapted to a multitude of atmospheric conditions.
\item It was developed to run on GPUs. The main advantage of using GPUs is the highly parallel architecture.
\end{itemize}
We have worked on a few upgrades to boost the performance of \texttt{THOR}, and the Venus-like simulations presented in this work using \texttt{OASIS} (including e.g., \texttt{THOR} and a full radiation scheme) take roughly 9 days using one NVIDIA V100 32GB graphic card. Other updates in the code are the inclusion of the radiative heating/cooling terms and wind friction from the physics modules directly into the dynamical equations (e.g., \citealt{2019Deitrick}). This update improves the stability of the model.   

\subsubsection{\texttt{CYCLOPS} - Radiative Transfer}
\texttt{CYCLOPS} is a fast radiative parametrization that represents the absorption/emission/scattering by gases and clouds. The radiative transfer scheme is the most computational expensive routine in \texttt{OASIS}. Our code is based on the work developed in \cite{2015Mendonca}. The stellar radiation  calculations are based on the delta-Eddington approximation (two-stream-type) with an adding layer method (multiple-scattering). For the thermal radiation case, the code is based on an absorptivity/emissivity formulation. More details on the equations solved can be found in \cite{2015Mendonca}. The formulation is now implemented to work on GPUs, which allowed us to make the code in \cite{2015Mendonca} more flexible and accurate. In the current code (\texttt{CYCLOPS}) we do not degrade the resolution of the cumulative probability function, which was done before to boost the performance of the radiative transfer code. Our radiation code is now fully integrated over 353 spectral bands and 20 Gaussian points, using the complete information of the original k-distribution table and avoiding any tunable parameters. The spectral resolution of the k-table changes as a function of spectral ranges, and we show those different resolutions in Table \ref{tab:spectral_res}. To boost the performance of the angular integration of the radiative transfer integration (emission solution), we have applied a 3-point Gaussian Quadrature integration instead of the typically used diffusivity factor. Our angular integrations increase the model flexibility and robustness.

The interaction of the radiation and air molecules depend largely on the molecular absorption cross sections. In our work, we have used the code \texttt{HELIOS-K} to calculate the molecular opacities (\citealt{2015Grimm}). We have applied a few changes to \texttt{HELIOS-K} to be able to produce opacities compatible with other works that have studied Venus (e.g., \citealt{2011Lee}): a) extended the code to include the CO$_2$ line shape suggested in \cite{1996Meadows}; b) used the data and methods from \cite{2017Gamache} to calculate the total internal partitions of the molecules; c) after the line-by-line information is gathered we use a MATLAB routine to tabulate the optical data using the methods from \cite{1991Lacis}. The molecules included in our study are CO$_2$, H$_2$O and SO$_2$, which are the main molecules that contribute to the radiative budget in the atmosphere. For further details about the absorption line parameters explored in this work please read \citealt{2011Lee}. The performance of our model was improved by building a unique k-table with the opacity of the three gases (CO$_2$, H$_2$O and SO$_2$) across different temperatures and pressures. The concentration of each gas in the atmosphere was kept constant with time (see module \texttt{MERLIN} below).

We added an absorption continuum to the line-by-line absorption to represent the collision-induced absorption and to alleviate the limitation due to the far-wing line truncation. The collision-induced absorption is generated in the deep atmosphere due to the collisions between the molecules that induce dipole moments and that can behave momentarily like more complex molecules. We implemented the same approach as suggested in \cite{2015Lebonnois}. The values to form the continuum were constrained from observations at fixed wavelengths: $0.7\times10^{-9}$ cm$^{-1}$ amg$^{-2}$ at 1.18 $\mu$m (\citealt{2011Bezard}), $5\times10^{-9}$ cm$^{-1}$ amg$^{-2}$ at 1.74 $\mu$m (\citealt{1995Bergh}) and $3.5\times10^{-8}$ cm$^{-1}$ amg$^{-2}$ at 2.4 $\mu$m (\citealt{2006Marcq}). As in \cite{2015Lebonnois}, we have interpolated the absorption continuum between the spectral windows referred to above. Due to the lack of observational or experimental constrains to the absorption continuum, the uncertainties above 2.3 $\mu$m are large. Above 3 $\mu$m we have added a base continuum of $3.0\times10^{-7}$ cm$^{-1}$ amg$^{-2}$ similar to the value used in \cite{2015Lebonnois}, which permits the model to calculate an average surface temperature of roughly 745 K. The collision-induced absorption values due to CO$_2$-CO$_2$ collisions at the wavelength range between 40 and 260 $\mu$m were taken from \cite{1997Gruszka} and \cite{2004Baranov}.

We consider the Rayleigh scattering from CO$_2$ molecules, which is associated with the elastic scattering of light by particles (in this case molecules) that are smaller than the wavelength of the radiation. Rayleigh scattering is inversely proportional to the fourth power of the wavelength, and in our work we followed the parametrization from \cite{2015Haus} to represent the scattering by CO$_2$ molecules.

As we explained in the introduction and will discuss in Section \ref{sec:clouds}, the cloud region in Venus contains an atmospheric constituent that causes a significant absorption of the stellar light at wavelengths between 0.32 and 0.8 $\mu$m. Despite the composition of the extra absorber is unknown, it is possible to constrain its spatial distribution and absorption efficiency from observations. We have implemented the same parametrization as in \cite{2015Haus}, which allowed us to include the UV absorber without specifying its composition. 

The cloud optical properties were calculated from Mie theory and made consistent with \cite{1986Crisp}. The composition of the cloud particles are assumed to be 25$\%$ water and 75$\%$ sulphuric acid, and the refractive indexes to calculate the optical properties of the clouds were taken from \cite{1975Palmer}. The surface reflectivity was set to 0.15 to represent the dark basaltic Venus' surface. You can read more about the Venus surface color in \cite{1986Pieters}.

The heating/cooling rates calculated from the radiative budget for each grid point have the following form:
\begin{equation}
Q = \frac{1}{\rho c_p}\frac{d F_{rad}}{dz}
\end{equation}
where $Q$ is the heating/cooling rates (K/s), $\rho$ the atmospheric density (kg/m$^3$), c$_p$ is the specific heat capacity (J kg$^{-1}$ K$^{-1}$), and F$_{rad}$ is the spectral integrated flux (W/m$^2$). F$_{rad}$ contains the flux contribution from both short-wave and long-wave radiative schemes that are updated every 55 simulated hours and 4 hours respectively. The quantity $Q$ is inserted in the entropy equation solved in \texttt{THOR}, which then calculates the new updated temperatures in the atmosphere.  

This code is very flexible and capable of exploring different atmospheric conditions and compositions. The main difficulty is to collect data that sufficiently well represents the optical properties at certain non Earth-centric conditions (e.g., collision-induced absorption of CO$_2$). 

The structure of \texttt{CYCLOPS} allows for a straightforward implementation of the dynamical-radiative-microphysical feedbacks on 3D simulations that will be explored soon with \texttt{OASIS} on Earth- and Mars-like atmospheres.

\begin{table}
\begin{tabular}{ |c|c| } 
\hline
Spectral range ($\mu$m) & Spectral resolution ($\mu$m)\\
\hline
1.70-2.55 & 0.02 \\ 
2.55-22.75 & 0.1 \\
40.75-100.75 & 10.0 \\
100.75-260.75 & 40.0 \\ 
\hline
\end{tabular}
\caption{Reference spectral resolution used in the 3D simulations.}
\label{tab:spectral_res}
\end{table}

\subsubsection{\texttt{ATLANTIS} - Surface and Soil}
\texttt{ATLANTIS} is the physics module in \texttt{OASIS} that represents the surface and soil of planets. A future implementation will also have a representation of the oceans. In the present work we have set the Venus surface to be a flat basalt surface. The evolution of the soil temperature in our Venus simulation was calculated using the following equation:
\begin{equation}
\frac{\partial T}{\partial t} = - \frac{\lambda}{C}\frac{\partial^2T}{\partial z_d^2}
\end{equation}
where $T$ corresponds to the different soil temperatures, $C$ is the specific heat per unit volume, $\lambda$ is the soil conductivity, and $z_d$ is the depth length from the surface. We have used 11 layers to represent the soil in the planet. The multilayer soil formulation used is similar to the scheme developed in \cite{1986Warrilow} and later used for the LMD Mars GCM (\citealt{1993Hourdin}) and Oxford Venus GCM (\citealt{2016Mendoncaa}). To represent the basalt soil thermal properties we have set the thermal inertia of the surface/soil to be 2200 J m$^{-2}$ K$^{-1}$ s$^{-\frac{1}{2}}$ (\citealt{1999Rees}). The formulation used allows us to also compute the radiative flux coming from the interior, which is then integrated in the energy budget of the surface to permit an accurate estimate of the surface temperature rate. 

\texttt{ATLANTIS} also includes the planetary boundary layer that represents the interaction between the atmosphere and surface. The development of \texttt{OASIS} is built on physical modules with different levels of complexity to facilitate gaining physical intuition about all the processes working in the simulations. The simulations examined in this work include very simple schemes to represent the boundary layer. The mechanical interaction between atmosphere and surface is represented in the reference simulations by a Rayleigh friction scheme identical to the formulation proposed in \cite{1994Held} to represent the Earth boundary layer in Earth simulations:
\begin{equation}
\frac{\partial \textbf{v}}{\partial t} = -k_v(\sigma) \textbf{v}
\end{equation}
where $\textbf{v}$ is the momentum and $\sigma$ is the ratio of the atmospheric pressure over the surface pressure. $k_v$ is the strength of the dissipation and it is modelled by:
\begin{equation}
k_v = k_f max\Big(0, \frac{\sigma-\sigma_b}{1-\sigma_b}\Big);
\end{equation}
where $\sigma_b$ is set to 0.7 and k$_f^{-1}$ is equal to 24 hours. For simplicity, in this work, we use the same set of parameters proposed to represent the Earth's surface friction in our Venus simulations. In Section \ref{sec:fri} we explore the impact of different dissipation strengths in the simulated atmosphere.

The excess of angular momentum observed in the atmosphere of Venus has to come from interchanges of angular momentum between the atmosphere and the surface (e.g., \citealt{2016Mendoncaa}). To improve the representation of this phenomenon in the simulations, we are testing more physically-based formulations based on bulk transport turbulent mixing parametrizations (e.g., \citealt{2016Mendoncaa}; \citealt{2017Read}) for future works. 

\subsubsection{\texttt{LOKI} - Turbulence}
In \texttt{LOKI}, we include a representation of the small-scale physics that cannot be resolved in the 3D GCMs, namely, turbulence. Various physical phenomena are difficult to resolve in space or time in 3D atmospheric simulations, because they require prohibitively large computational resources. An alternative option is to recover their impact on the simulations by parametrizing their physical behaviour in the large-scale phenomena. 

Convection occurs in the atmosphere when the temperature gradient becomes steeper than an adiabatic profile. The convectively formed instabilities have an important role mixing, for example, heat and momentum in the atmosphere. There is a diversity of parametrizations that have been implemented in climate models to represent the small-scale convection with different levels of complexity. In this work we have implemented a simple routine that does not require any tuning. We have implemented a simple convective adjustment routine that has been used in other climate models, such as the LMD GCM (\citealt{1993Hourdin}). Our very efficient  parametrization mixes the enthalpy instantaneously in a buoyant unstable atmospheric column. The equation solved is:
\begin{equation}
	\label{equ:mixpt}
	\bar{\theta} =\frac{\int^{p_{top}}_{p_{bot}} C_p\theta \Pi d p}{\int^{p_{top}}_{p_{bot}} C_p\Pi dp}
\end{equation}
where $\Pi$ is the exener function ($(p/p_0)^{Rd/Cp}$), and $p_{top}$ and $p_{bot}$ are the pressures at the top and bottom of the unstable column, and $\theta$ the potential temperature\footnote{The potential temperature is the temperature that a parcel of dry air would attain if brought adiabatically from its initial state to a standard pressure (usually the surface pressure).} (e.g., \citealt{2018Mendoncab}). Note that in this parametrization the only physical quantity that is mixed is potential temperature; other quantities such as momentum remain fixed. The convection scheme is largely dependent on heat capacity, C$_p$, and we explore its impact on the simulation in Section \ref{sec:cp}.

Included in \texttt{LOKI}, we have a ``sponge layer'' that prevents spurious wave reflections in the top rigid boundary that can compromise the stability of the numerical simulations and more importantly that can lead to erroneous results in the top most layers of the atmosphere. We followed the same formulation as explored in \cite{2018Mendoncab}. The scheme is formulated so that it attempts to represent wave breaking phenomenon in the atmosphere. We have formulated the code to linearly damp the eddy components of the momentum and temperature in the upper part of the atmosphere. The forcing increases with altitude and is limited to a shallow region in the top model domain. The formulation used also makes use of a very efficient method developed in \cite{2018Mendoncab} to compute the eddy component of the wind from an icosahedral grid.

An essential numerical scheme to keep the atmospheric models stable is the numerical dissipation. Every model needs some type of dissipation to avoid the accumulation of high frequency waves at the smallest scale resolved by the numerical model, mimicking the physical cascade of energy to smaller scales. The turbulence and eddy viscosity in the sub-grid scale is represented by using a fourth-order hyperdiffusion operator coupled with a 3D divergence damping. The formulation is the same as used in \cite{2016Mendoncab}. We have also extended the original version from \cite{2016Mendoncab}, by implementing the vertical component of the divergence damping to make the divergence operator more homogeneous in 3D. However, this extra component did not have any impact on the results due to the still crude horizontal resolution compared to the vertical (the divergence damping strength scales with the spatial resolution). For more information on the technical details and implementation please see \cite{2016Mendoncab}.

\subsection{Prescription for gas composition and clouds}
In this work we use simple representations of the clouds and chemistry in Venus. Clouds play a very important role in the Venus' atmospheric circulation and climate. The cloud structure remains constant during the simulations. The cloud properties were constrained by Venus observations and are divided into three different particle size modes (\citealt{1980Knollenberg} and \citealt{1986Crisp}): mode 1 corresponds to a mean particle size less than 0.4 $\mu$m and extend from 32 to about 90 km altitude; mode 2 particles have a mean size of 1 $ \mu$m and form part of the main cloud deck from roughly 50 km to 80 km altitude; the largest particles are included in the mode 3 with a mean particle radius of 3.65 $\mu$m and located between 50 and 60 km. We have included the composition of the four main gases in the atmosphere. The concentration of the different gases were assumed to be well mixed in the atmosphere and close to the defined in the Venus International Reference Atmosphere (VIRA, \citealt{1985Zahn}). The CO$_2$ concentration is set to 0.965 vmr\footnote{vmr - volume mixing ration.},  H$_2$O to 5$\times 10^{-5}$ vmr, SO$_2$ to 1$\times 10^{-4}$ vmr and the rest in N$_2$. The concentrations are not meant to be exactly equal to the values observed but to capture the main bulk conditions of a Venus-like planet.

\section{Reference Simulations}
\label{sec:ref_sim}

The reference simulation tests include a complete coupling of \texttt{THOR} (dynamics), \texttt{CYCLOPS} (radiation), \texttt{LOKI} (turbulence) and \texttt{ATLANTIS} (surface and soil), with a constant cloud cover and chemical composition.

\begin{figure*}
\begin{centering}
%\subfigure[4$^o$ - 10,000 days]{
%\includegraphics[width=0.5\columnwidth]{u_psi_ref_gl4_1}}
%\subfigure[4$^o$ - 15,000 days]{
%\includegraphics[width=0.65\columnwidth]{u_psi_ref_gl4_2}}
%\subfigure[4$^o$ - 20,000 days]{
%\includegraphics[width=0.65\columnwidth]{u_psi_ref_gl4_3}}
\subfigure[4$^o$ - 25,000 days]{
\includegraphics[width=1.0\columnwidth]{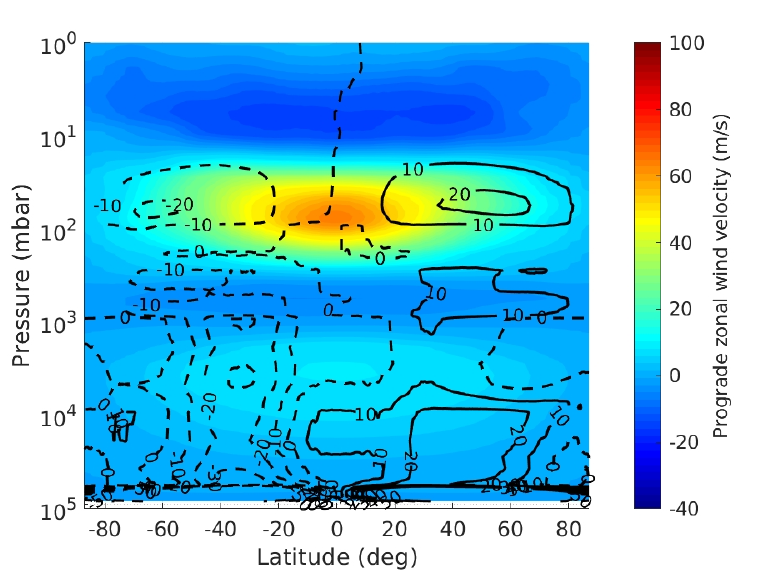}\label{fig:Ref_U-a}}
%\\
%\subfigure[2$^o$ - 10000 days]{
%\includegraphics[width=0.5\columnwidth]{u_psi_ref_gl5_1}}
%\subfigure[2$^o$ - 15,000 days]{
%\includegraphics[width=0.65\columnwidth]{u_psi_ref_gl5_2}}
%\subfigure[2$^o$ - 20,000 days]{
%\includegraphics[width=0.65\columnwidth]{u_psi_ref_gl5_3}}
\subfigure[2$^o$ - 25,000 days]{
\includegraphics[width=1.0\columnwidth]{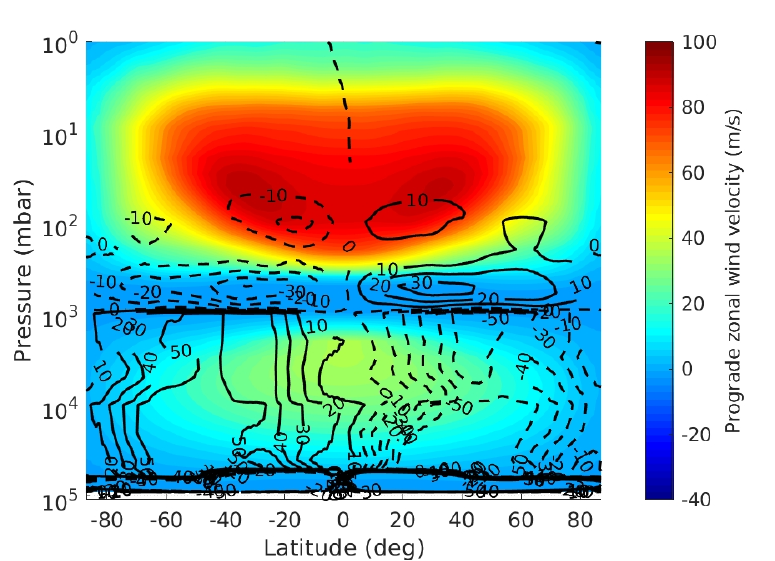}}
\caption{Zonally and timed averaged zonal winds and mass stream function for the reference simulation with different spatial resolution: (a) 4 degrees  and (b) 2 degrees. The units of the zonal winds are in m$/$s and the positive numbers point in the prograde direction of the planet. The mass stream function is shown as the contours and is in units of 10$^{10}$ kg$/$s. The results were averaged over 1,000 Earth days.}
\label{fig:Ref_U}
\end{centering}
\end{figure*}

\subsection{MIRAGE}
The input to our simulations is shown in Table \ref{tab:model}, and the composition of the atmosphere and surface, and the cloud parameters as described in the previous Sections. Before \texttt{OASIS} starts running the 3D computations, a module called \texttt{MIRAGE} calculates the 1D radiative-convective solution from the set of planet parameters chosen by the modeller. The module \texttt{MIRAGE} combines the radiative transfer scheme with the convective adjustments and the soil codes. The \texttt{MIRAGE} computations are performed in a column of the 3D grid, which is only discretized vertically in altitude. The stellar flux is globally averaged according to \cite{1989Crisp}, and the simulations are integrated until the temperature in each grid cell converges to a steady value (radiative-convective solution). The global-averaged temperature-pressure profile obtained with the 1D MIRAGE module is later used to initialize the temperatures and pressures of every column in the 3D simulation (all the columns start with the same input). This procedure allows us to speed-up the convergence of the deep thermal structure in Venus-like planets where the radiative timescales are hundreds of Earth days. 

The module \texttt{MIRAGE} is also used to test new physical modules before implementing the 3D configuration, allow fast exploration of the planetary parameter space, and build a physical intuition on complex planetary climate problems. \texttt{MIRAGE} can reproduce the emission and transmission spectra from the 1D radiative-convective temperature profiles but those capabilities are not explored in this work.

\subsection{3D Reference Simulations}
\label{sec:3dsimu}
\texttt{OASIS} has integrated the Venus simulations for 25,000 Earth days. The initial temperature in each column was set to the temperature-pressure profiles obtained from the \texttt{MIRAGE} module (as explained in the previous Section), and the initial wind speed is set to a solid body rotation with an equatorial maximum of 30 m$/$s. The initial wind structure helps the model to start from a more stable solution than starting from rest. It is important to note that it has been shown that the spin-up phase of the Venus simulations are roughly 25,000 Earth days (e.g., \citealt{2007Lee3}), which means that by the end of the simulations the initial states have been forgotten and the climate solutions converge to a statistical steady state. For the reference 3D simulations we have run two simulations that started from the same initial conditions but with different spatial resolutions: roughly 4 degrees and 2 degrees. The two simulations used exactly the same input, however, the time-step of the 2 degree resolution was reduced to 50 seconds to satisfy stability criteria from the dynamical core \texttt{THOR}. Fig. \ref{fig:Ref_U} shows the zonal winds (longitudinal component of the wind field) and mass stream function at the end of the simulations. 

The 4 degrees model has 4 times fewer grid columns than the model with 2 degrees, which results in a simulation roughly 4 times faster. The 4 degrees simulation has obtained qualitatively the same climate features as the 2 degrees simulation, however, quantitatively the magnitude of the equatorial jets and the meridional circulation are distinct. We find that the inability of the 4 degree spatial resolution simulations in producing the strong winds is associated with the numerical errors in the advection scheme that become comparable to the physical sources/sinks in the deep atmosphere (see for example \citealt{2012Lebonnois} or \citealt{2008Thuburn} for a more detailed description on the conservation properties in dynamical cores). The amplitude of this error will vary for different methods used in the dynamical cores, which means that the threshold in spatial resolution is expected to vary for different models. However, despite the differences, both simulations obtain similar wind and temperature distributions. Both simulations with different spatial resolution show two regions of local maxima in the winds: at pressure levels near the cloud top ($\sim$100 mbar) and in the lower atmosphere at pressures between 1 and 10 bars. In both resolution experiments the model produced two shallow direct circulation cells within the 10 lowest kilometres. In the 2 degrees simulation, the mass stream function shows, above this region the formation of indirect cells that extend from the shallow direct cells to roughly 1 bar (roughly the pressure level of the cloud base in the simulations). The atmospheric circulation in the 4 degrees simulation is more complex in the region between 10 and 1 bar: the zonal winds are weaker than the 2 degrees simulations, and the mass stream function shows a complex pattern that evolves with time, similar to the results obtained in \cite{2016Mendoncab}. However, the atmospheric cells in this region tend to form large extended direct cells. Above the cloud base the atmosphere is largely driven by the absorbed stellar radiation, and both simulations form large Hadley cells.

\begin{figure}
\begin{centering}
\subfigure[4 degrees simulation]{\label{fig:Ref_Temp_lon1}
\includegraphics[width=1.0\columnwidth]{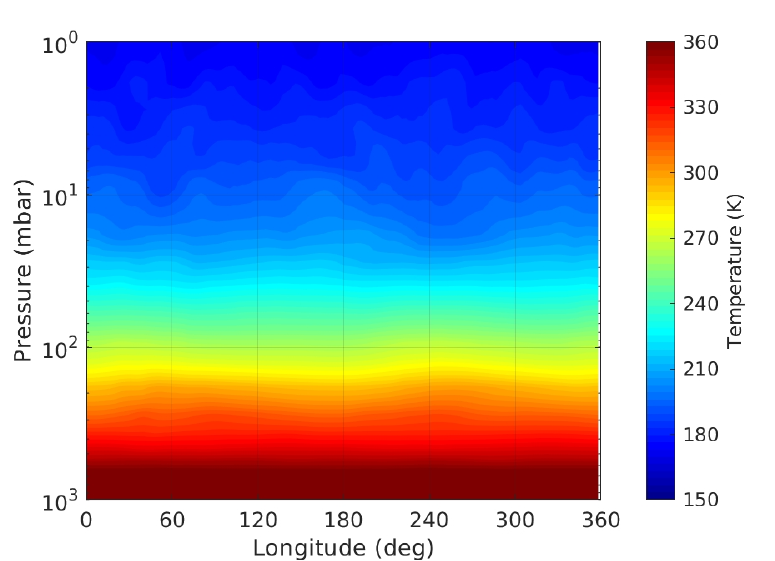}}
\subfigure[2 degrees simulation]{\label{fig:Ref_Temp_lon2}
\includegraphics[width=1.0\columnwidth]{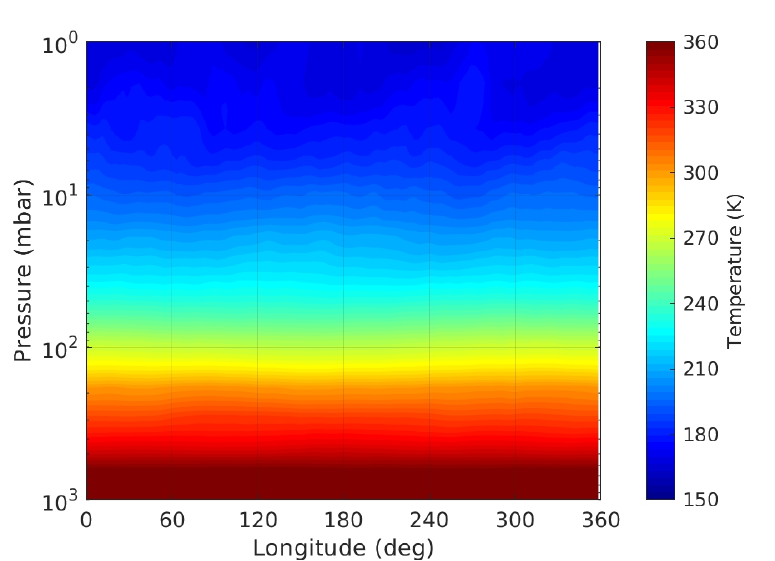}}
\label{fig:Ref_T}
\caption{Temperature maps of the upper atmosphere along the equatorial region obtained at the end of the simulations with 4 degrees (a) and 2 degrees (b). The values were averaged in latitude between -20$^o$ and 20$^o$.}
\label{fig:Ref_Temp_lon}
\end{centering}
\end{figure}

\begin{figure}
\begin{centering}
\subfigure[Global mean temperatures]{\label{fig:Ref_Temp1}
\includegraphics[width=1.0\columnwidth]{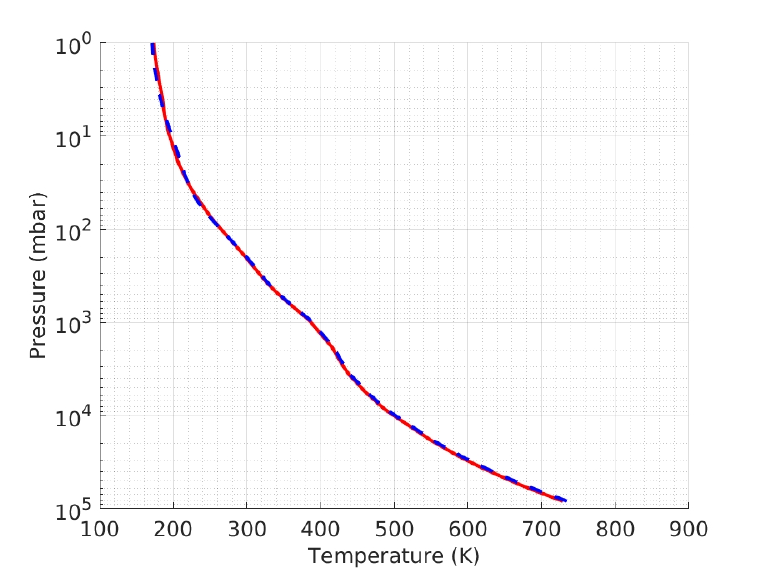}}
\subfigure[4 degrees simulation]{\label{fig:Ref_Temp2}
\includegraphics[width=1.0\columnwidth]{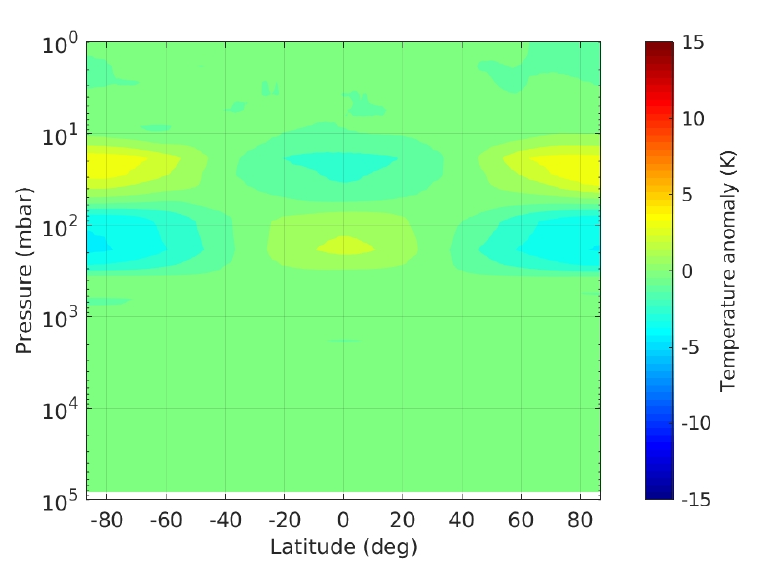}}
\subfigure[2 degrees simulation]{\label{fig:Ref_Temp3}
\includegraphics[width=1.0\columnwidth]{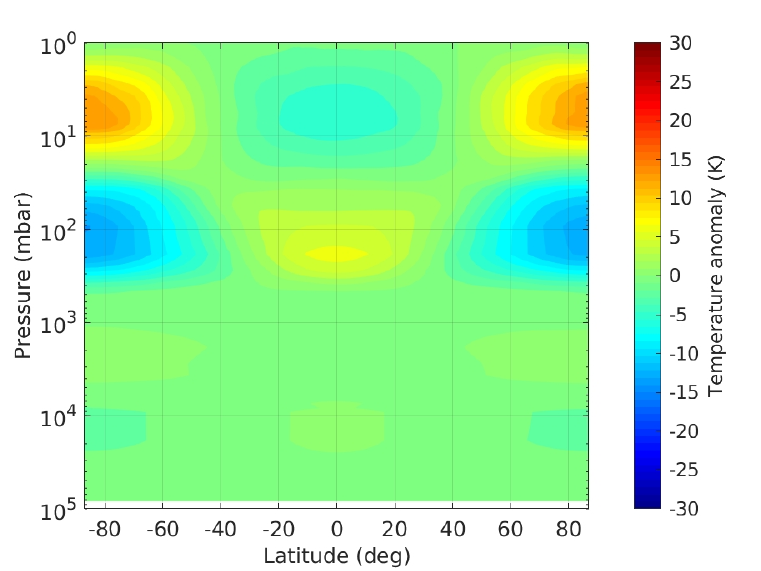}}
\label{fig:Ref_T}
\caption{(a) map of the horizontally and time (1,000 Earth days) averaged temperatures. The two lines represent the two simulations with different space resolution: the red solid line is the 4 degrees simulations and the blue dashed line the 2 degrees. (b) and (c) are the zonal and time averaged (over 1,000 Earth days) temperature anomaly maps (see equation \ref{eq:tempano}).}
\label{fig:Ref_Temp}
\end{centering}
\end{figure}

The global mean temperature for the reference simulations are shown in Fig. \ref{fig:Ref_Temp1}. The global mean temperature-pressure profiles for the simulations with different spatial resolutions are very similar, with a surface temperature of roughly 745 K. The temperature differences along the longitude in the upper atmosphere (above the pressure level 1 bar) are small as it shown in Fig. \ref{fig:Ref_Temp_lon}. The upper atmosphere is the region where the temperature differences are expected to be larger due to the small radiative timescales and also because it is the region where most of the stellar radiation is absorbed. The small differences are associated with the efficient heat transport by the winds from the day to the night side. As we will see in Section \ref{sec:obs}, due to the small differences in temperature along the longitude, the differences between the infrared spectra from day and night side are also very small. The amplitude of the diurnal and semi-diurnal thermal tides in the atmosphere above the clouds are a few Kelvin, and similar to the results found in other GCMs such as \cite{2016Mendoncaa} and \cite{2016Lebonnois}.

The strong winds overlaying a slowly rotating planet set Venus on a cyclostrophic regime (e.g., \citealt{2012Mendonca}). In this regime the centrifugal acceleration balances the geopotential gradient term (\citealt{1973Leovy}), which links the temperature structure with the atmospheric circulation. In this approximation a positive latitudinal gradient of the temperature is associated with a negative vertical gradient of the zonal winds. To clearly show the temperature structure as a function of latitude, in Figs. \ref{fig:Ref_Temp2} and \ref{fig:Ref_Temp3} we show the temperature anomaly ($T_a$) maps. The anomalies were calculated from the following equation:
\begin{equation}
\label{eq:tempano}
T_a = T - \frac{\int_{\lambda=0}^{\lambda=2\pi} \int_{\phi=-\pi/2}^{\phi=\pi/2} T A^2\cos\phi d\lambda d\phi}{\int_{\lambda=0}^{\lambda=2\pi} \int_{\phi=-\pi/2}^{\phi=\pi/2} A^2\cos\phi d\lambda d\phi}
\end{equation}
where $T$ is the absolute temperature, $A$ is the radius of the planet, $\phi$ is the latitude and $\lambda$ is the longitude. The hotter poles are produced by the compressional adiabatic heating caused by the downward branch of the large Hadley cells. The equator-to-pole differences are larger in the simulation with 2 degrees resolution, which is associated with the stronger winds formed in this simulation. The temperature differences in the 2 degree simulation are between 20 to 30 K at 100 mbar, which is compatible with the observational work from Venus Express that shows equator-to-pole temperature differences of 30 K \citep{2009Tellmann}. However, at 1 bar the results based on observations \citep{2009Tellmann} also show equator-to-pole temperature differences of 30 K, with a latitudinal gradient in the temperature starting at high latitudes (roughly 60 degrees latitude), which is not reproduced in our simulations. Other Venus GCMs that used non-gray radiative transfer such as, \cite{2010Lebonnois}, \cite{2016Mendoncaa} and \cite{2016Lebonnois}, also underestimate the equator-to-pole temperature differences at 1 bar, which could be related with the poor representation of the cloud distribution across the atmosphere in the models. The 4 degree simulation obtained similar temperature anomaly pattern than the 2 degree simulation, however, the winds that have half of the absolute magnitude, produced equator-to-pole differences with roughly half of the temperature anomalies observed in the 2 degree simulation.

Despite the differences in the magnitudes between the simulations with 2 and 4 degrees, both produce qualitative the same atmospheric circulation and temperature structure. The simulation with 4 degrees is also similar to the simulations reproduced by  \cite{2010Lebonnois} and \cite{2016Mendoncaa} that used similar physical modules but different dynamical core based on finite difference methods and a space resolution of 5 degrees. Since we can capture roughly the same temperature structure and wind distribution of the 2 degree simulations with the 4 degree simulation, we decided to explore the impact of the clouds, different heat capacity of the atmosphere and surface friction in the simulated atmospheres with the 4 degree setup because simulation is 4x faster than the 2 degree simulation. 

\section{Comparison with observations}
\label{sec:obs}

The goal of this work was not to interpret new detailed data of Venus, or reproduce accurately the current Venus state, but to simulate the broad climate and circulation features that characterize Venus. However, as it is shown below, \texttt{OASIS} obtains robust results that reproduce quantitatively the temperature structure and circulation in Venus. 

\begin{figure}
\begin{centering}
\includegraphics[width=1.0\columnwidth]{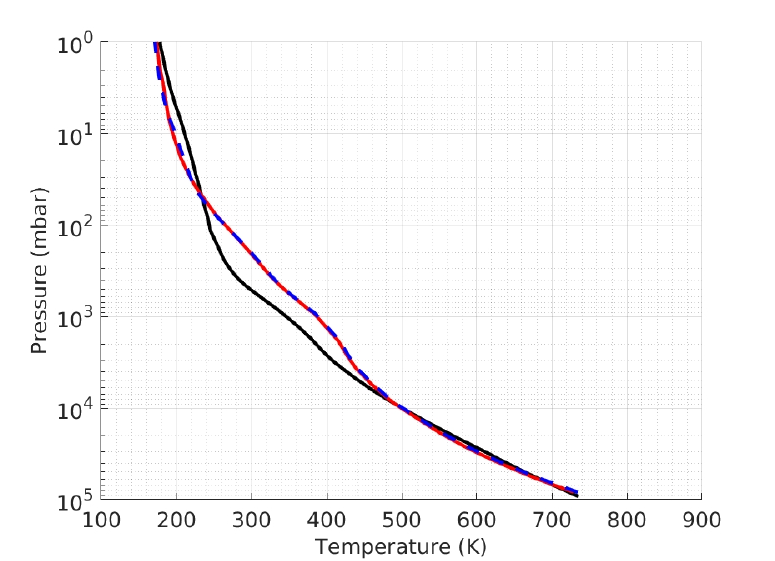}
\caption{Global mean temperature from observations and \texttt{OASIS}. The solid black line is the VIRA (Venus International Reference Atmosphere) temperature profile based on observations of Venus (\citealt{1985Kliore}). The \texttt{OASIS} profiles overlap each other: blue dashed line (2 degrees simulation) and red solid line (4 degrees simulation).}
\label{fig:temp_G_models}
\end{centering}
\end{figure}

The global mean temperature from observations and the \texttt{OASIS} simulations are shown in Fig. \ref{fig:temp_G_models}. As we saw in the previous Section the two simulations with different resolution obtain almost identical global mean temperature profiles. The observational profile is based on the results presented in \cite{1985Kliore} (Venus International Reference atmosphere). There is good agreement between the model and observations, however, in the cloud region, the model results are consistently warmer than the observations. The temperature in this region is sensitive to the cloud distribution and parameters (\citealt{2015Lebonnois}), and we could have reduced the differences between the model results and the observations by tuning those parameters. However, finding the correct cloud distribution and parameters to match exactly the Venus current state goes beyond the scope of this paper. Our results are, for example, consistent with \cite{2011Lee} that used the same cloud distribution as we did (from \citealt{1986Crisp}) and a more computationally expensive radiative transfer scheme from the model DISORT (\citealt{2000Stamnes}). 

\begin{figure}
\begin{centering}
\includegraphics[width=1.0\columnwidth]{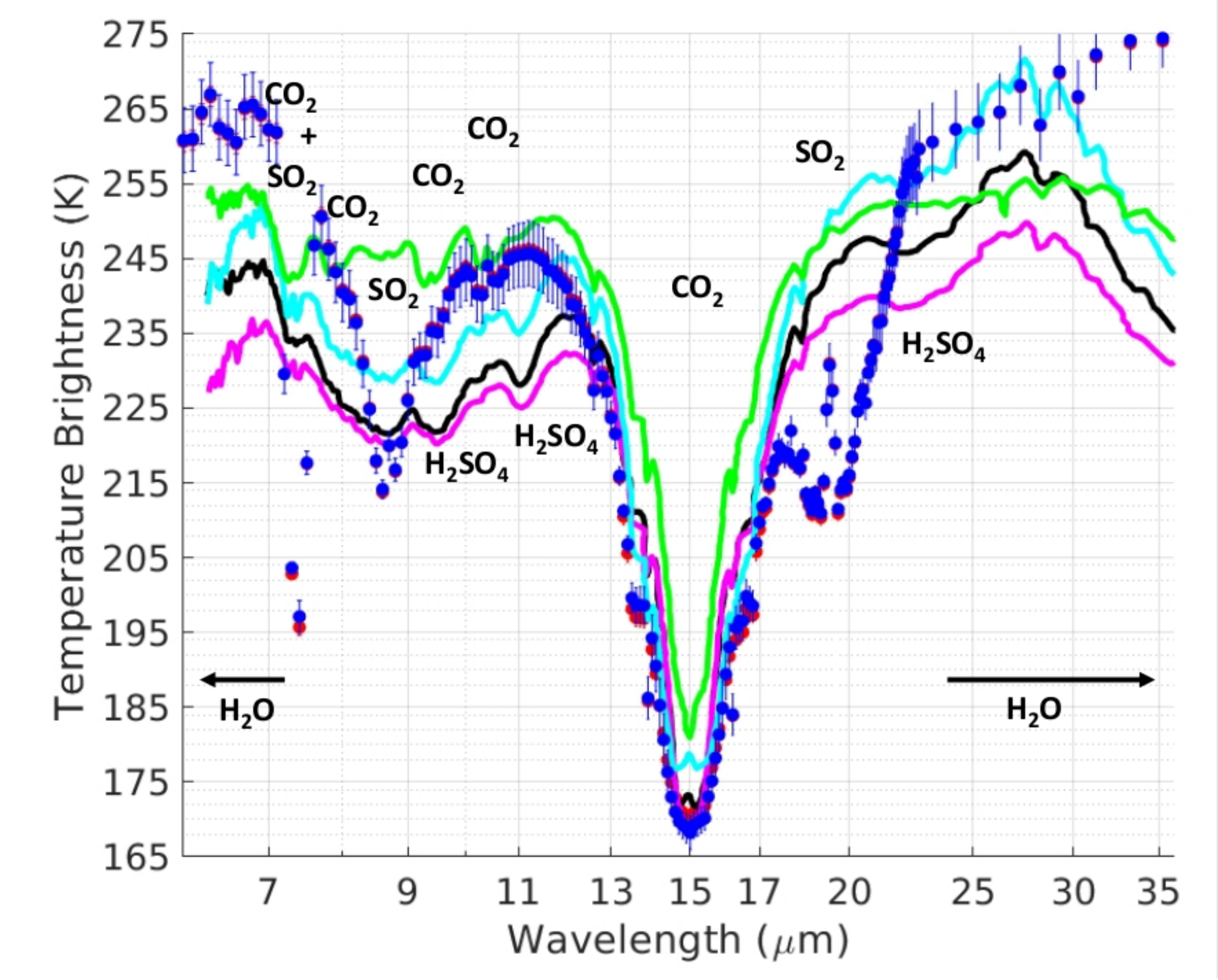}
\caption{Brightness Temperature from \texttt{OASIS} and observations from Venera 15 (\citealt{2004Zasova}). The solid lines are results from observations at different regions in the atmosphere (\citealt{2004Zasova}): (black solid line) latitude < 35$^o$, L$_s$ = 20-90$^o$; (magenta solid line) latitude < 35$^o$, L$_s$ = 270-310$^o$; (cyan solid line) 10$^o$ < latitude < +10$^o$, L$_s$ = 75$^o$; (green solid line) North polar region, latitude > 85$^o$; where L$_s$ is the solar longitude. The main absorption features from CO$_2$, H$_2$O, SO$_2$ and H$_2$SO$_4$ are indicated in the figure. The blue and red points are results from \texttt{OASIS} with a space resolution of 2 and 4 degrees respectively. The model results correspond to globally averaged values of the brightness temperature for different wavelengths. The uncertainties of the model values is the standard deviation of the globally averaged values.}
\label{fig:tb_obs}
\end{centering}
\end{figure}

Fig. \ref{fig:tb_obs} shows the brightness temperature as a function of wavelength from \texttt{OASIS} and observations of Venus. The model results presented in Fig. \ref{fig:tb_obs} are at the same spectral resolution than the spectral resolution used in the \texttt{OASIS} simulations. The observational values were obtained from different missions and also at different locations in the atmosphere as explained in the caption of the figure. The brightness temperature is a measure of the upward infrared radiance at the top of the atmosphere, and can provide information about the vertical temperature structure and composition of the atmosphere. There is a good agreement between the observed values and the OASIS results close to the center of the CO$_2$ lines, for example, at 15 $\mu$m. The good representation of the model CO$_2$ line is associated with the temperatures of the upper atmosphere, above the cloud deck, being very close to the observed values. Note that there is almost no difference between the simulations with different resolution. The main differences between the model results and the observations are at the wavelengths of the SO$_2$ absorption lines, such as, at around 7, 9 and 20 $\mu$m. In these regions the brightness temperature from \texttt{OASIS} is consistently lower than the values observed. The differences with respect to the observations around the SO$_2$ features is expected since we have assumed a well-mixed SO$_2$ concentration in the atmosphere that is higher than the values observed in the Venus upper atmosphere (e.g., \citealt{2018Taylor}). The brightness temperature observed in Venus around the SO$_2$ lines corresponds to radiation coming mostly from the upper cloud region. On the other hand, by increasing the levels of concentration of SO$_2$ in our model, we see that the regions above the clouds, where the temperatures are colder, have a larger contribution to the brightness temperature. The higher values of SO$_2$ absorption in the upper atmosphere could be used in the future to explore signs of volcanic activity in other planetary atmospheres. The water features represent emission from the upper cloud region, and as we saw before in Fig. \ref{fig:temp_G_models}, the temperatures in the cloud region in our reference simulation are warmer than the observed values. 

\begin{figure}
\begin{centering}
\includegraphics[width=1.0\columnwidth]{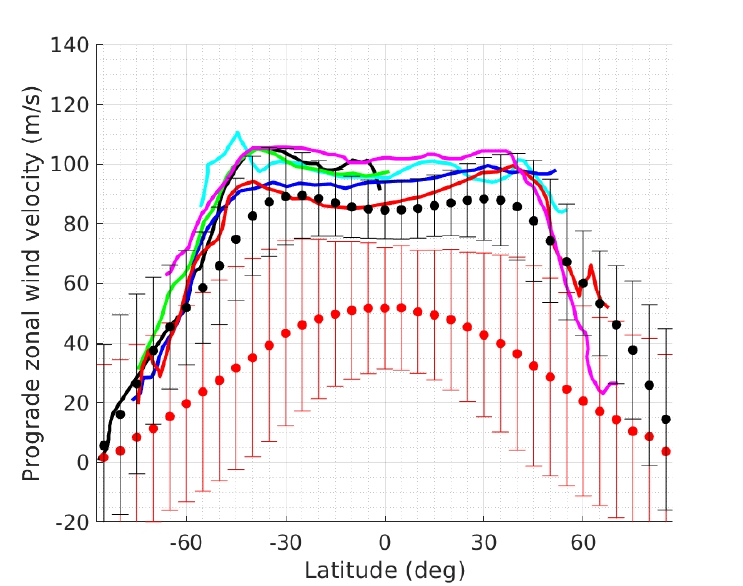}
\caption{Zonal wind profiles from observations and the \texttt{OASIS} simulations. The winds plotted correspond to an altitude range of 65-70 km. The observational values were estimated by tracking the cloud motions at UV wavelengths in multiple space missions (\citealt{2017Lavega}): Mariner 10, 1974 (cyan solid line); Pioneer-Venus, 1980 (blue solid line); Pioneer-Venus, 1982 (red solid line); Galileo, 1990 (magenta solid line); Venus Express VIRTIS, 2006–2012 (black solid line); Venus Express VMC, 2006–2012 (green solid line). The filled circles correspond to the model results: red (simulation with 4 degrees spatial resolution) and black (simulation with 2 degrees spatial resolution). The model results were spatially averaged over the day-side hemisphere to be consistent with the UV observations and the uncertainties correspond to the standard deviation. The results were time averaged over 1,000 Earth days.}
\label{fig:u_obs}
\end{centering}
\end{figure}

Venus is known to exhibit very strong atmospheric winds. Several different observational methods have been used to study the atmospheric circulation in Venus, however, the circulation below the cloud base continues poorly constrained observationally (e.g., \citealt{2017Lavega}). A very common method to measure the wind speed in Venus is based on cloud tracking of multiple temporal images. The cloud tracking method assumes that the cloud components behave as passive tracers of the atmospheric flow and allows to estimate the magnitude and direction of the horizontal winds. There is a large amount of data from cloud tracking measurements of the Venus atmosphere (e.g., \citealt{2017Lavega}) and we compare our model results with the observations from multiple space missions in Fig. \ref{fig:u_obs}. The results of the model and the observed values show clouds moving in Venus at a velocity around 100 m$/$s, which is roughly 60x faster than the rotational velocity of the solid planet. The simulation with 4 degrees resolution obtained winds consistently weaker than the observations, and a large variability on the magnitude as evident from the large standard deviation. Notice that, as explained in section \ref{sec:3dsimu}, the main cause for the difference in the magnitude of the winds is the numerical precision in the advection scheme, which is improved when using better spatial resolution. The simulation with 2 degrees reproduced strong winds that are largely consistent with the observations within the standard deviation. Despite the slightly weaker winds in the equatorial region, the model results represent the Venus circulation in the cloud region very well. This represents a great success for the \texttt{OASIS} platform since 3D GCM simulations of the Venus circulation are very challenging (\citealt{2017Lavega}).

\section{Sensitivity Experiments}
\label{sec:sen_tests}
The experiments in this Section were integrated with a model configuration similar to the 4 degrees reference simulation described in Section \ref{sec:ref_sim}.

\subsection{Clouds}
\label{sec:clouds}
The clouds in Venus have an important impact on the planet's energy budget and in its appearance. The presence of the clouds in the atmosphere causes a large fraction of the solar energy to be deposited in the cloud region, which has a strong influence on the waves excited in that region that drive the atmospheric circulation (e.g., \citealt{2010Lebonnois} and \citealt{2016Mendoncaa}). As explained in the introduction, the clouds in Venus are a mixture of sulphuric acid H$_2$SO$_4$ and small amounts of H$_2$O (typically 25$\%$). In the cloud region, we also find a UV absorber which still has, as discussed in the introduction, an unknown composition. In this Section, we attempt to determine the impact of the unknown UV absorber and the clouds in the temperature and atmospheric circulation. Studying the case of a cloud-free Venus atmosphere is also interesting because Venus may have experienced periods of cloud free events. The main component of the clouds, H$_2$SO$_4$, is produced photochemically from the photolysis of SO$_2$ and CO$_2$ (e.g., \citealt{2017Catling}). However, the abundances of SO$_2$ and H$_2$O are driven by the exospheric escape of hydrogen and heterogeneous reactions with surface minerals and volcanic outgassing. In \cite{2001Bullock}, it is suggested that the Venus clouds would largely dissipate on a time-scale of 30 million years without continued volcanic out-gassing.

\begin{figure}
\begin{centering}
\subfigure[Cloud experiments]{
\includegraphics[width=1.0\columnwidth]{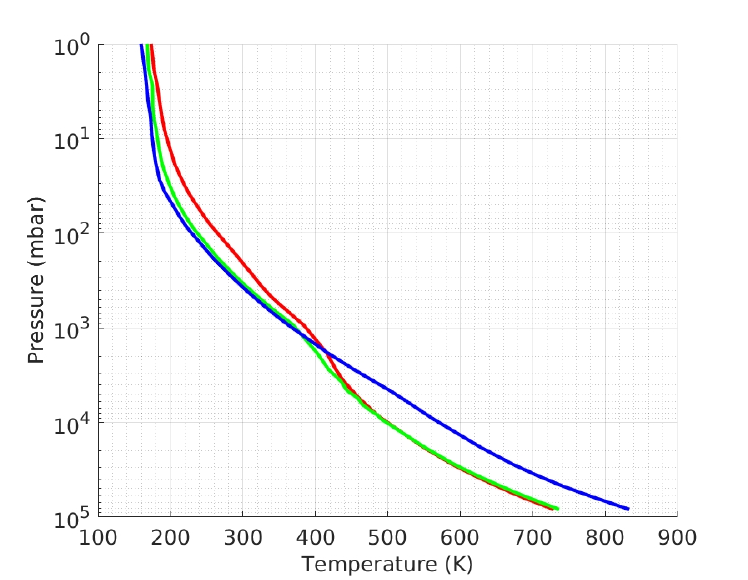}\label{fig:Ref_Temp_CLDS-a}}
\subfigure[No extra UV absorption]{
\includegraphics[width=1.0\columnwidth]{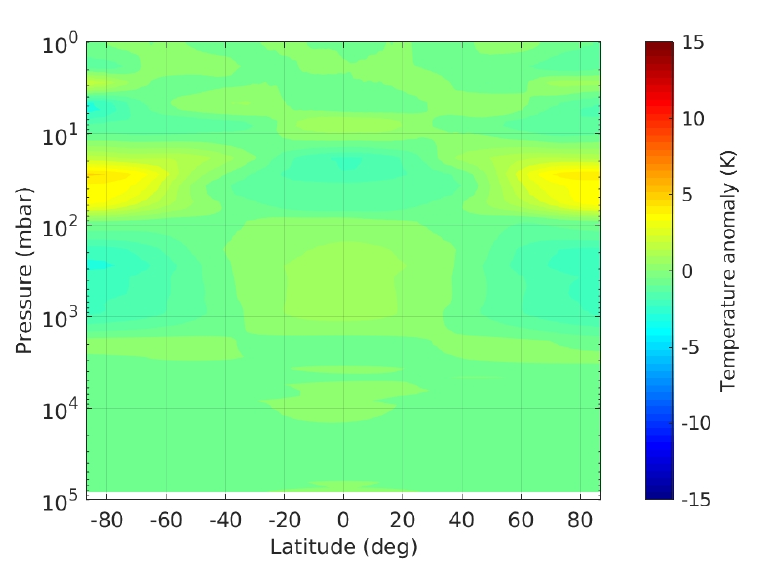}\label{fig:Ref_Temp_CLDS-b}}
\subfigure[No clouds or UV absorber]{
\includegraphics[width=1.0\columnwidth]{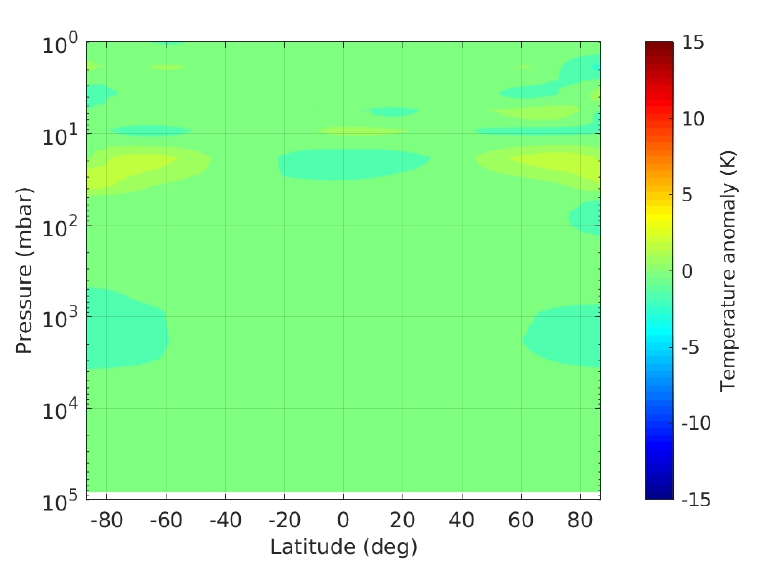}\label{fig:Ref_Temp_CLDS-c}}
\caption{Temperatures obtained in the cloud experiments. (a) map of the horizontally and time (1,000 Earth days) averaged temperatures. The three coloured solid lines represent the three different experiments: green, no extra UV absorption; red, reference simulation; blue, no clouds or extra UV absorption. (b) and (c) are the zonal and time averaged (over 1,000 Earth days) temperature anomaly maps (see Eq. \ref{eq:tempano}).}
\label{fig:Ref_Temp_CLDS}
\end{centering}
\end{figure}

\begin{figure}
\begin{centering}
\subfigure[No extra UV absorption]{
\includegraphics[width=1.0\columnwidth]{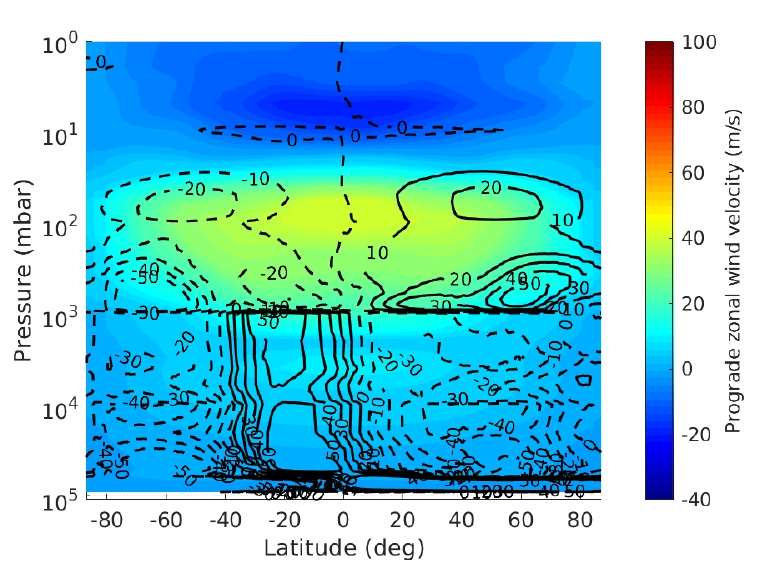}\label{fig:U_CLDS-a}}
\subfigure[No clouds and no extra UV absorption]{
\includegraphics[width=1.0\columnwidth]{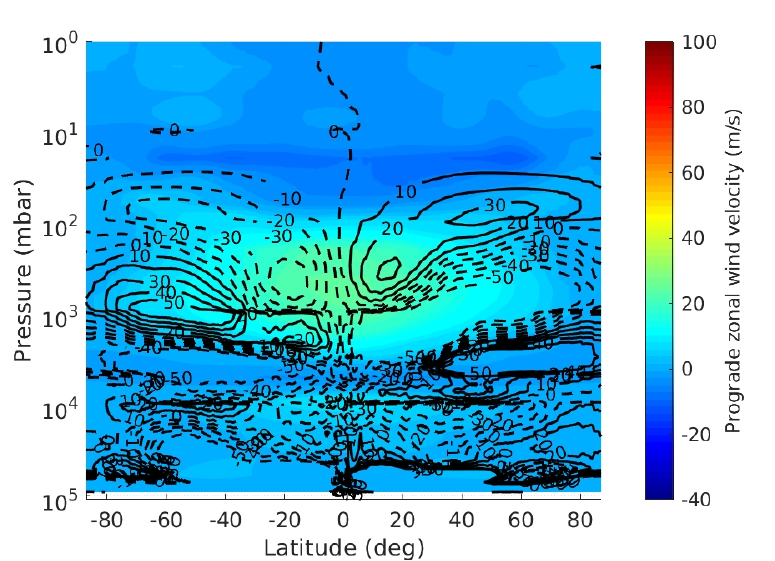}\label{fig:U_CLDS-b}}
\caption{Zonally and time averaged zonal winds and mass stream function for cloud experiments: (a) no extra UV absorption and (b) no clouds and no extra UV absorption. Both simulations use 4 degrees spatial resolution and the results were averaged over 1,000 Earth days. The units of the zonal winds are in m$/$s and the positive numbers point in the prograde direction of the planet. The mass stream function is shown as the contours and is in units of 10$^{10}$ kg$/$s.}
\label{fig:U_CLDS}
\end{centering}
\end{figure}

We run two simulations: one with H$_2$SO$_4$ clouds but no UV absorber, and another with no UV absorber or clouds. In Fig. \ref{fig:Ref_Temp_CLDS}, we investigate the main impact of the clouds in the temperature structure of Venus from our 3D simulations. Fig. \ref{fig:Ref_Temp_CLDS-a} shows the global mean temperatures. The reference simulation described in the previous Section is the red line. If the unknown UV absorber is removed from the simulations (green solid line in Fig. \ref{fig:Ref_Temp_CLDS-a}) it affects the temperatures of the upper atmosphere, which becomes colder. The deep atmosphere is only slightly affected by a very small heating of the planet surface. If the clouds are completely removed (including the unknown UV absorber) it has a large effect in the temperature: the temperature in the upper atmosphere becomes similar to the case of just removing the unknown UV absorber (blue solid line in Fig. \ref{fig:Ref_Temp_CLDS-a}), which is an indication that the unknown UV absorber has a larger impact on the temperatures of the upper atmosphere than the H$_2$SO$_4$/H$_2$O clouds; the lower atmosphere becomes much hotter, which is because the energy that was being reflected by the clouds is now reaching deeper regions in the atmosphere (including the surface). With no clouds, the atmosphere becomes considerably less opaque to the solar radiation in the 1-0.01 bar region, and scattering becomes less important to the atmospheric radiative budget. In this case, the atmospheric structure produced becomes similar to results found with a gray thermal radiative transfer (\citealt{2012Robinson}) where the atmosphere develops a deep dry convective region that extends from the surface to roughly 0.1 bar. Figs. \ref{fig:Ref_Temp_CLDS-b} and \ref{fig:Ref_Temp_CLDS-c}, show the temperature anomaly maps from the two cloud experiments. If we remove just the unknown UV absorber, the amplitude of the temperature anomalies are weaker than the reference simulation, but their spatial distribution is similar. Removing the clouds completely has a large impact, and the anomalies become very small (Fig. \ref{fig:Ref_Temp_CLDS-c}). Fig. \ref{fig:U_CLDS} shows the winds produced by the cloud experiments. As expected the winds produced by the case with no unknown UV absorber (Fig. \ref{fig:U_CLDS-a}) is similar to the reference simulations, but produces weaker zonal winds. It is also worth noting that the atmospheric direct cells in the cloud region become stronger. The upper atmosphere becomes more transparent to the UV radiation causing a larger latitudinal gradient of the radiative heating in the lower cloud region and that induces the formation of stronger atmospheric cells. With no clouds (Fig. \ref{fig:U_CLDS-b}) the zonal winds become very weak with a stronger meridional wind component. More energy is deposited in the deep atmosphere, which creates a more dynamically active lower atmosphere: multiple strong atmospheric cells are stacked on the top of each other. The stronger vertical mixing shown in the simulation with no clouds may have an important impact on the chemical spatial distribution for planets with similar characteristics. See the work of \cite{2018Lincowski} for a discussion on the impact of the clouds in the 1D radiative-convective equilibrium temperature profile for a planet with Venus-like composition around the TRAPPIST-1 star.

\subsection{Atmospheric heat capacity}
\label{sec:cp}
A constant heat capacity of the atmosphere is a good approximation for the Earth. However, in the massive Venus CO$_2$ atmosphere, the heat capacity has a strong dependence on the temperature, and can change from roughly 1200 J$/$K$/$kg at the surface to 600 J$/$K$/$kg at 100 km altitude. Other Venus GCMs have included the dependence of the heat capacity on the temperature (e.g., \citealt{2010Lebonnois} and \citealt{2016Mendoncaa}). To include the dependence of the heat capacity on the temperature in a GCM requires that some important quantities such as the potential temperature are re-formulated. As shown in \cite{2010Lebonnois} and \cite{2016Mendoncaa}, the dynamical core of the GCMs (the part that solves the fluid equations) has to be modified to take this new formulation of potential temperature into account. More details on the implementation of a variable heat capacity in GCMs can be found in \cite{2010Lebonnois} and \cite{2016Mendoncaa}. Other changes are easier to implement, such as the calculation of the radiative heating rates and convective adjustment. In \cite{2016Mendoncaa}, the entropy was mixed in the atmosphere in a superadiabatic atmospheric column instead of the usually used enthalpy (in the subsection \texttt{LOKI} we describe the convective adjustment routine used in the current version of \texttt{OASIS}). In this work, we have decided to fix the value of the heat capacity. To test the impact of using different heat capacity values in the simulations we run two more simulations: heat capacity of 700 and 1100 J K$^{-1}$kg$^{-1}$.

\begin{figure}
\begin{centering}
\subfigure[Heat capacity experiments]{
\includegraphics[width=1.0\columnwidth]{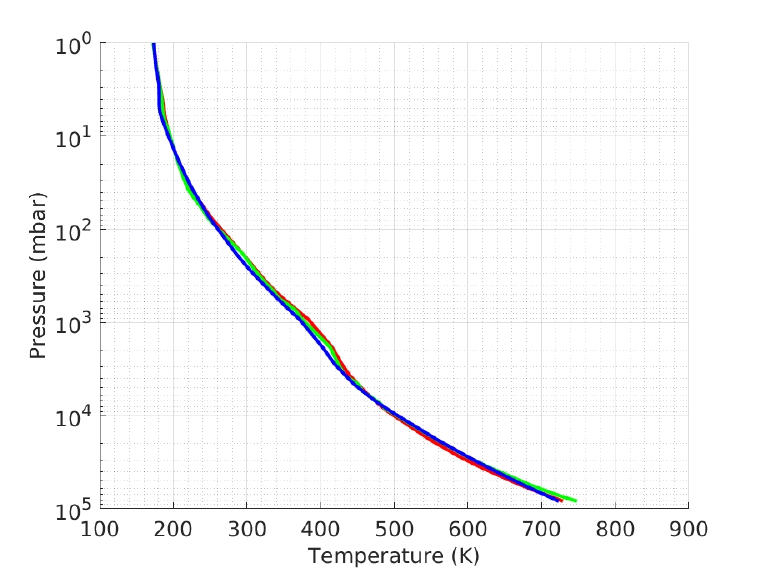}
\label{fig:Ref_Temp_CP-a}}
\subfigure[Heat capacity equals to 700 J K$^{-1}$kg$^{-1}$]{
\includegraphics[width=1.0\columnwidth]{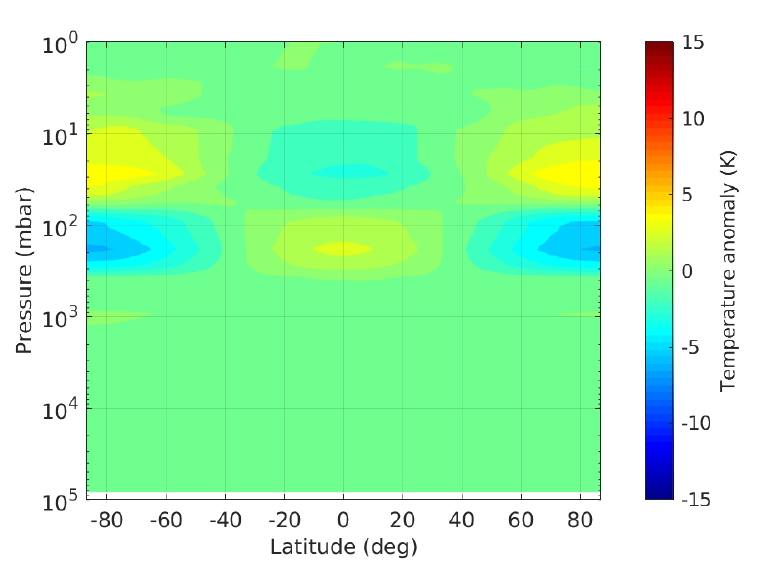}\label{fig:Ref_Temp_CP-b}}
\subfigure[Heat capacity equals to 1100 J K$^{-1}$kg$^{-1}$]{
\includegraphics[width=1.0\columnwidth]{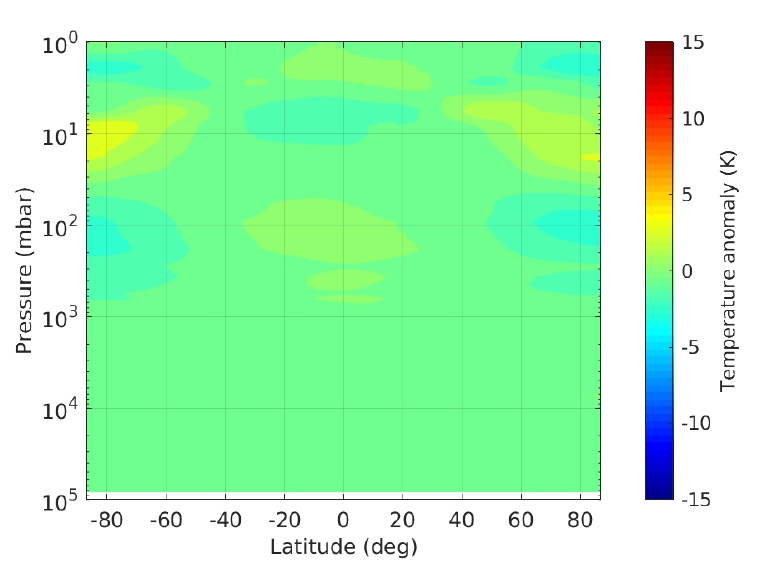}
\label{fig:Ref_Temp_CP-c}}
\caption{Temperatures obtained in the heat capacity experiments. (a) map of the horizontally and time (1,000 Earth days) averaged temperatures. The three coloured solid lines represent the three different heat capacity values used: green, 700 J K$^{-1}$kg$^{-1}$; red, 900 J K$^{-1}$kg$^{-1}$; blue, 1100 J K$^{-1}$kg$^{-1}$. (b) and (c) are the zonal and time averaged (over 1,000 Earth days) temperature anomaly for heat capacity of 700 J K$^{-1}$kg$^{-1}$ and 1100 J K$^{-1}$kg$^{-1}$ respectively.}
\label{fig:Ref_Temp_CP}
\end{centering}
\end{figure}

Fig. \ref{fig:Ref_Temp_CP-a} shows the horizontal and time averaged temperatures of the three simulations: two with the different heat capacity values and the reference simulation. The profiles look very similar. The largest differences are located in the deepest atmosphere and the cloud region. There is almost no difference between the case with 900 and 1100 J K$^{-1}$kg$^{-1}$ in the hot deep atmosphere, however, the case with 700 J K$^{-1}$kg$^{-1}$ obtained a hotter surface. The hotter values are associated with less energy being transported vertically by convection, which allows the formation of a steeper lapse rate. Figs. \ref{fig:Ref_Temp_CP-b} and \ref{fig:Ref_Temp_CP-c} show the temperature anomaly maps that are similar to the reference simulations. However, the anomalies in the simulation using 1100 J K$^{-1}$kg$^{-1}$ are weaker than in the other experiments. 

\begin{figure}
\begin{centering}
\subfigure[Heat capacity equals to 700 J K$^{-1}$kg$^{-1}$]{
\includegraphics[width=1.0\columnwidth]{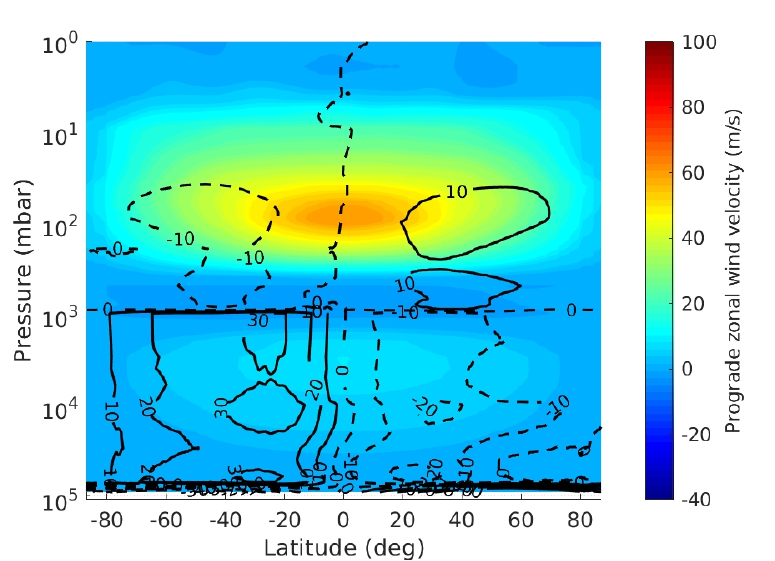}\label{fig:U_CP-a}}
\subfigure[Heat capacity equals to 1100 J K$^{-1}$kg$^{-1}$]{
\includegraphics[width=1.0\columnwidth]{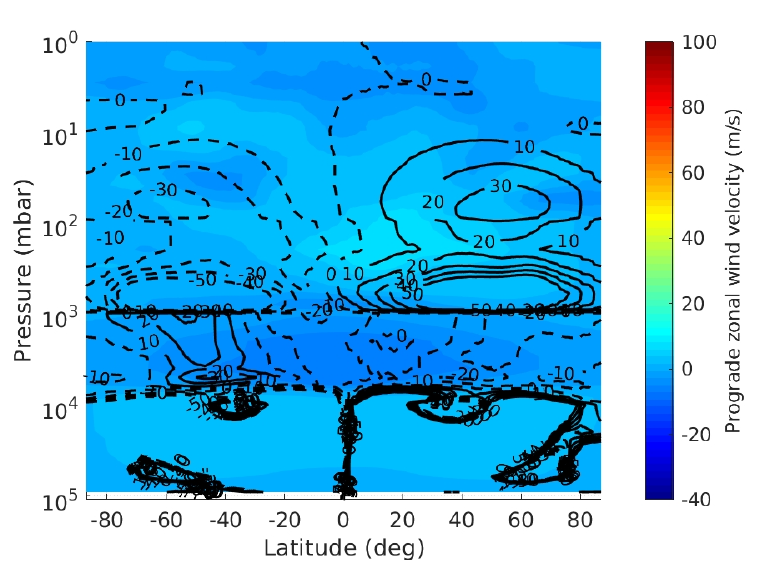}\label{fig:U_CP-b}}
\label{fig:U_CP}
\caption{Zonally and timed averaged zonal winds and mass stream function for simulations with different heat capacity values: (a) 700 and (b) 1100 J K$^{-1}$kg$^{-1}$. Both simulations use 4 degrees spatial resolution and the results were averaged over 1,000 Earth days. The units of the zonal winds are in m$/$s and the positive numbers point in the prograde direction of the planet. The mass stream function is shown as the contours and is in units of 10$^{10}$ kg$/$s.}
\end{centering}
\end{figure}

The winds produced in these experiments are shown in Figs. \ref{fig:U_CP-a} and \ref{fig:U_CP-b}. The wind pattern and strength in the simulation with heat capacity of 700 J K$^{-1}$kg$^{-1}$ are very similar to the results obtained in the reference simulations. On the other hand, the results when using a larger heat capacity produced wind speeds very different from the reference simulation. The mass stream function in Fig. \ref{fig:U_CP-b}, shows an atmosphere very different from the reference simulation at deep pressures. The difference is caused by a more convectively active atmosphere in the case of the simulation with 1100 J K$^{-1}$kg$^{-1}$. The larger heat capacity allows the mixing in the atmosphere to be extended to larger atmospheric columns, which drags the horizontal motion. The mechanism accelerating the equator is a combination of zonal mean circulation, thermal tides and transient waves (\citealt{2015Mendonca}), which is similar to the mechanism driving the atmospheric circulation in hot Jupiter planets (see for example \citealt{2017Mayne}). As shown in \cite{2017Zhang}, increasing the heat capacity can reduce the wave speed in the atmosphere, and disrupts the formation of a strong jet. 

\subsection{Surface friction}
\label{sec:fri}
The physical properties of the planet's surface will be challenging to constrain through observations. However, the boundary layer has an important role representing a boundary condition that regulates, for example, the exchanges of heat and momentum between the surface and the atmosphere. 

\texttt{OASIS} uses a very simplified boundary layer scheme as explained in Section \ref{sec:OASIS}, which linearly drags the winds in the boundary layer region. This parametrization is a very simple representation of the mechanical interaction between the surface and the atmosphere and it is controlled by a parameter that sets the strength of the surface friction. In this experiment we reduce the strength of the friction and study its impact on the simulated atmosphere. 

\begin{figure}
\begin{centering}
\subfigure[Low surface friction experiment - Horizontally and time averaged temperatures]{
\includegraphics[width=1.0\columnwidth]{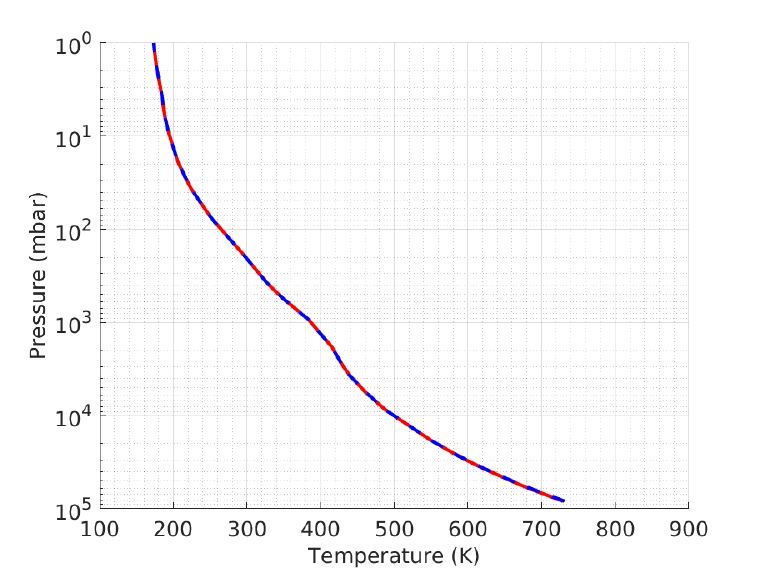}\label{fig:Ref_Temp_LOWFRI-a}}
\subfigure[Temperature anomaly maps]{
\includegraphics[width=1.0\columnwidth]{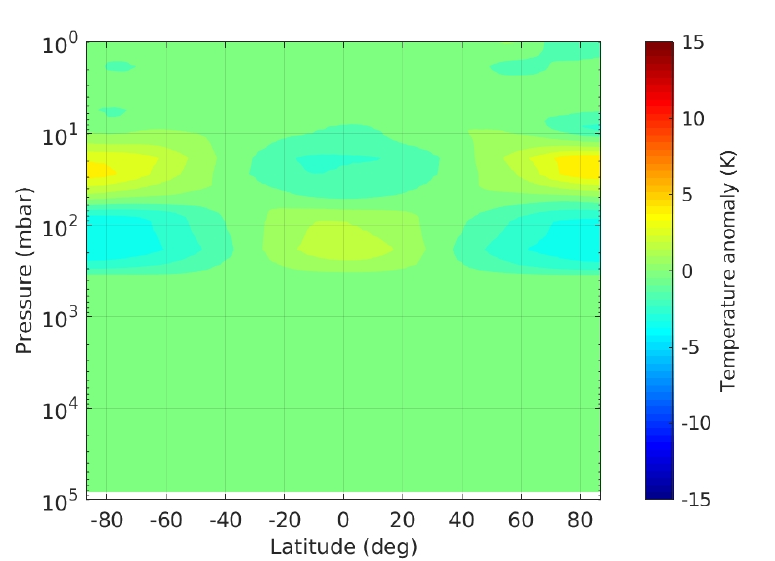}\label{fig:Ref_Temp_LOWFRI-b}}
\caption{Temperatures obtained in the surface friction experiment (k$_f^{-1}=10$ Earth day). (a) map of the horizontally and time (1,000 Earth days) averaged temperatures. The two colors correspond to the reference simulation with 4 degree spatial resolution (solid red line) and the low surface friction simulation (blue dashed line). (b) is the zonal and time averaged (over 1,000 Earth days) temperature anomaly maps.}
\label{fig:Ref_Temp_LOWFRI}
\end{centering}
\end{figure}

\begin{figure}
\begin{centering}
\includegraphics[width=1.0\columnwidth]{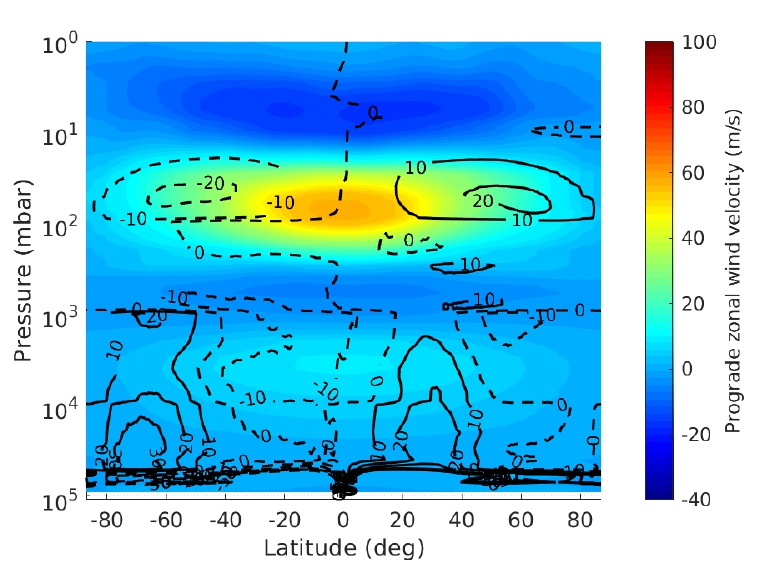}
\caption{Zonally and timed averaged zonal winds and mass stream function for simulation with different lower surface friction ($k_f^{-1} = 10$ Earth days). This simulation uses 4 degrees spatial resolution and the results averaged over 1,000 Earth days. The units of the zonal winds are in m$/$s and the positive numbers point in the prograde direction of the planet. The mass stream function is in units of 10$^{10}$ kg$/$s}
\label{fig:U_LOWFRI}
\end{centering}
\end{figure}

Figs. \ref{fig:Ref_Temp_LOWFRI} and \ref{fig:U_LOWFRI} show the temperature and wind maps from the low friction experiment (surface friction 10x weaker than the reference simulation). The results obtained are almost identical to the reference simulation. The winds produced have a very important role driving in the temperature structure in Venus, like the temperature anomalies in latitude and small temperature differences in longitude. Since the lower surface friction does not have an important impact on the atmospheric winds produced, the temperatures obtained in this experiment are similar to the ones in the reference simulations.

\section{Venus seen as an exoplanet}
\label{sec:venus_exoplanet}
\texttt{OASIS} is a theoretical tool capable of modelling the environment of planets and it can also be used to plan observations or build new ideas for observational methods. In this Section, we explore our available tools to study the predicted spectral observations if the simulated planet of the previous Sections (i.e., a Venus-like planet) were to be observed at a distance of 10 pc from Earth. Considering the characteristics of the planet-star system, we have focused on exploring the emission and reflected spectra of the planet. As we will see in the results below, a Venus-like planet around a Sun-like star will be possible to characterize with an observational facility such as the LUVOIR mission concept (\citealt{2018LUVOIR}). Our goal in this section is not to prove that a Venus-like planet can be observed by the LUVOIR mission concept, which has already been shown in \cite{2018LUVOIR}, but to show that the new \texttt{OASIS} modules produce the expected observed spectra. Nevertheless, our results, as we show below are consistent with the results from \cite{2018LUVOIR}.

\begin{figure}
\begin{centering}
\subfigure[Reference simulation]{
\includegraphics[width=1.0\columnwidth]{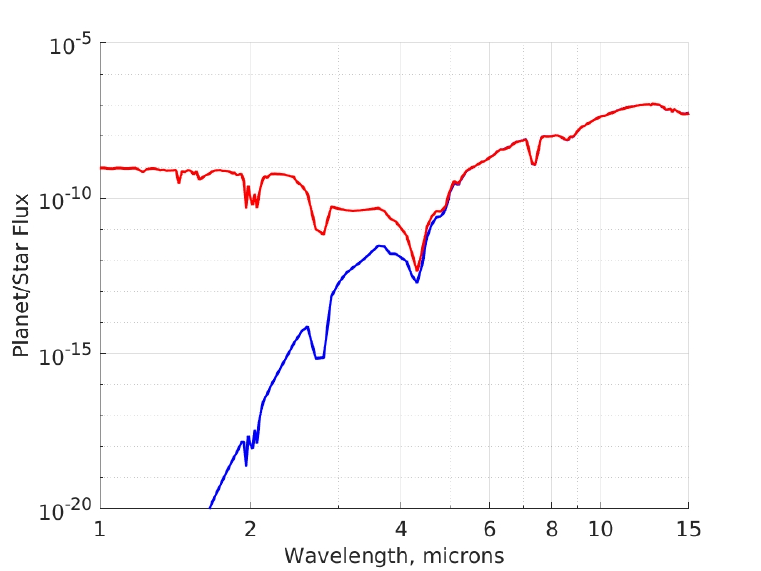}\label{fig:spec_planet-a}}
\subfigure[Experiment with no clouds or extra UV absorption]{
\includegraphics[width=1.0\columnwidth]{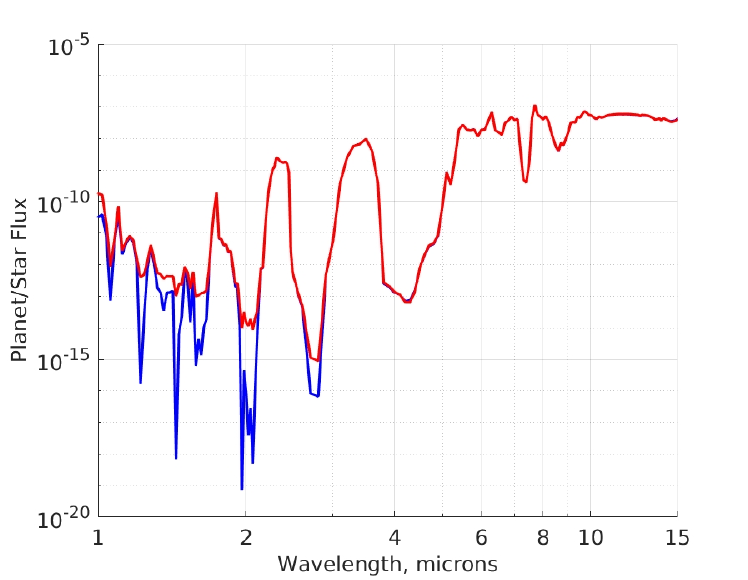}\label{fig:spec_planet-b}}
\caption{Flux ratios for two simulations: (a) reference simulation and (b) experiment with no clouds and no extra UV absorption. The red lines correspond to the planet fluxes from the day-side of the planet and the blue lines to the night-side.}
\label{fig:spec_planet}
\end{centering}
\end{figure}

In Fig. \ref{fig:spec_planet}, we present the fluxes coming from the simulated planet, which include the emitted planet flux and the stellar radiation being reflected in the planet (obtained from the 3D simulations). The spectral resolution used in the figure is the same as the resolution used in the 3D simulations. In the figure, the two panels represent different simulations: (a) refers to the reference simulation; (b) refers to the simulation with no clouds and the unknown UV absorber. The coloured solid lines represent the flux coming from the day side (red) and night side (blue) of the planet. In the reference simulation the impact of the clouds is very clear in the results shorter than 4 $\mu$m. The clouds are very reflective and help to significantly raise the outgoing flux. These results suggest that for a massive atmosphere such as the one in Venus, a highly reflective cloud structure helps raise the number of photons coming from the planet for wavelengths shorter than 4 $\mu$m, which otherwise would be absorbed in the deep atmosphere. In Fig. \ref{fig:spec_planet-a}, there is a large contrast between the day and night sides of the planet, however, the differences decrease for wavelengths longer than 4 $\mu$m. At longer wavelengths the fluxes originate mostly from the upper atmosphere (above the clouds) where the atmospheric circulation has redistributed efficiently the heat and reduced the differences in temperature between the day- and night-side of the planet. In Section \ref{sec:ref_sim}, we saw that the differences in temperature between the day and night side are relatively small. In Fig. \ref{fig:spec_planet-b}, the spectrum shows stronger spectral signatures than in Fig. \ref{fig:spec_planet-a}. These stronger signatures correspond to information from the lower atmosphere, which is now exposed due to the absence of the cloud structure. The fluxes are also higher due to the hotter deep atmosphere plus surface. The differences between the day-night side in the deep atmosphere (Fig. \ref{fig:spec_planet-b}) are mostly due to the differences in temperature between the two sides of the planet. The differences in temperature are caused by the winds being less efficient mixing the heat in the deep atmosphere compared with the upper atmosphere. The contribution from the surface albedo (0.15) to the planet flux is very small.

\begin{figure}
\begin{centering}
\subfigure[Reference simulation]{
\includegraphics[width=1.0\columnwidth]{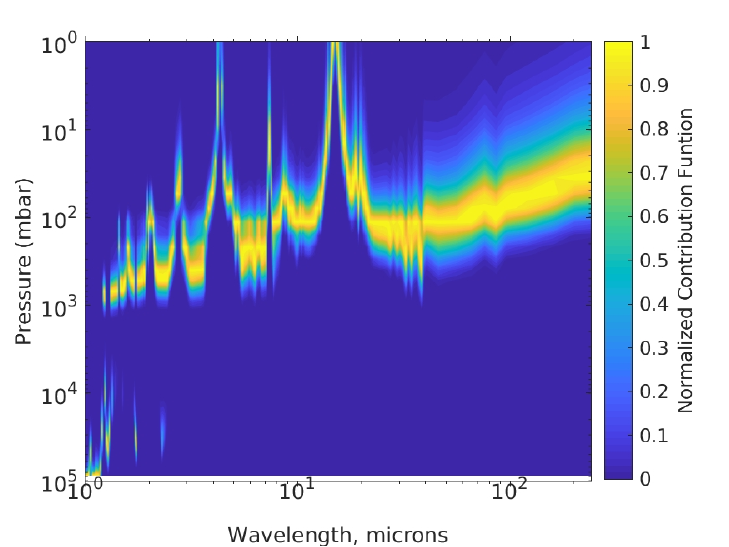}}
\subfigure[Cloud free experiment]{
\includegraphics[width=1.0\columnwidth]{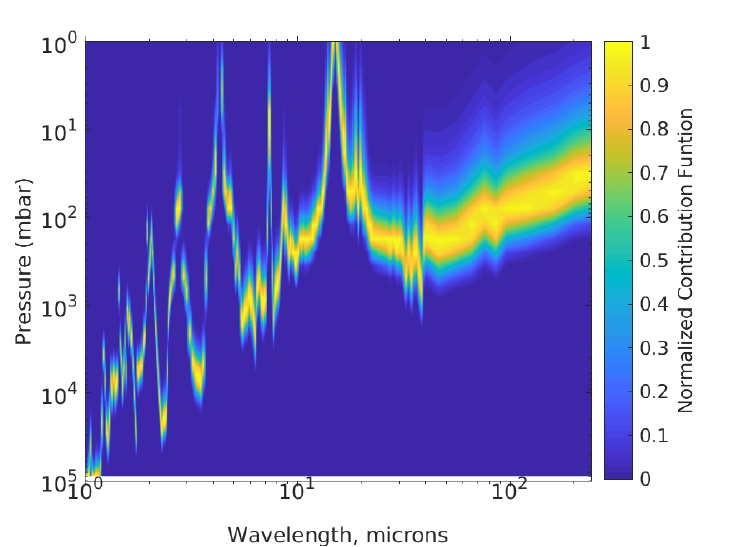}}
\caption{Contribution function as a function of wavelength for two simulations: reference and with no clouds.}
\label{fig:Contri_F}
\end{centering}
\end{figure}

The fluxes at different wavelength in Fig. \ref{fig:spec_planet} contain information from different regions in the atmosphere. A good diagnostic to explore from which pressure range the planet emission flux emerges is the contribution function. The contribution function (C$_f$) has the following form (\citealt{2009Knutson}), 
\begin{equation}
C_f(p,\lambda) = B(\lambda, T) e^{-\tau}\frac{d \tau}{d\log(p)},
\end{equation}
where the function $B$ is the Planck function, $T$ is the absolute temperature, $\tau$ is the optical depth and $p$ is the pressure. Fig. \ref{fig:Contri_F}, shows the contribution function for the reference simulation and the simulation with no clouds. As can be seen from the two panels, the cloud deck has an important impact shaping the contribution function. The main impact of the clouds is the block of the upward radiation coming from the deep layers for wavelengths shorter than 10 $\mu$m. Most of the emitted outgoing radiation at wavelengths shorter than 10 $\mu$m comes from the cloud deck at 1-0.1 bars. Note that the GCM resolves some spectral windows at wavelengths near 1 and 2 $\mu$m. Looking at these spectral windows we can probe the extreme climate conditions of the deep atmosphere in the cloudy Venus (e.g., \citealt{1984Allen}). The peak shape features in the contribution function maps are related to the main absorption features of CO$_2$, for example at roughly: 1.6, 2, 2.7, 4.3 and 15 $\mu$m. For wavelengths longer than 10 $\mu$m the emitted flux contributions are from pressure levels above the cloud base.

\begin{table}
\begin{center}
\caption{Instrument parameters used to simulate the possible spectrum obtained with LUVOIR. The formulation is based on \citealt{2016Robinson}.}
\begin{tabular}{ | l | l |}
\hline
 Parameters & Value  \\ \hline \hline
 Star Radius ($R_s$) & 1.0 \\ \hline
 Number of exozodis & 1.0 \\ \hline 
 Distance (pc) & 10.0 \\ \hline 
 Wavelength minimum ($\mu$m) & 0.4 \\ \hline 
 Wavelength maximum ($\mu$m) & 2.5 \\ \hline  
 Integration time (hours) & 10 \\ \hline 
 Outer working angle ($\lambda/$D) & 20.0 \\ \hline 
 Raw contrast & 10$^{-10}$ \\ \hline
 Instrument spectral resolution ($\lambda/\Delta\lambda$) & 70 \\ \hline
 Telescope and instrument throughput & 0.2 \\ \hline
 Dark current ($s^{-1}$) & $1\times10^{-4}$ \\ \hline
 Horizontal pixel spread of IFS spectrum & 3 \\ \hline 
 Read noise per pixel & 0.1 \\ \hline
 Size of photometric aperture ($\lambda/$D) & 1.5 \\ \hline
 Quantum efficiency & 0.85 \\ \hline
 V-band zodiacal light \\ surface brightness (mag arcsec$^{-2}\mu$m$^-1$) & 23 \\ \hline 
 V-band exozodiacal light \\ surface brightness (mag arcsec$^{-2}\mu$m$^-1$) & 22 \\ \hline
 Detector maximum exposure time (hours) & 1 \\ \hline
 Telescope mirror temperature (K) & 80 \\ \hline
 Effective emissivity for the \\ observing system (of order unity) & 0.9 \\ \hline
 \label{tab:model_LUVOIR}
 \end{tabular}
\end{center}
\end{table}

\begin{figure}
\begin{centering}
\subfigure[Telescope diameter, 8 meters]{
\includegraphics[width=1.0\columnwidth]{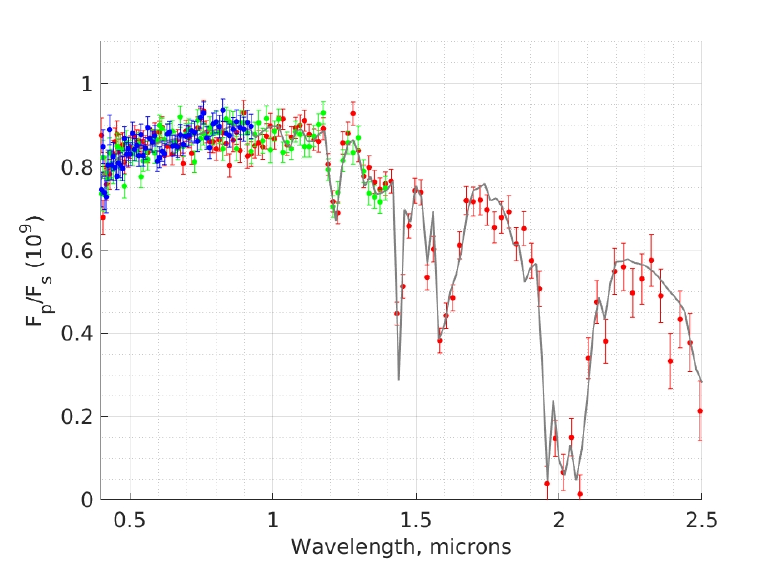}\label{fig:Venus_LUVOIR-a}}
\subfigure[Telescope diameter, 15 meters]{
\includegraphics[width=1.0\columnwidth]{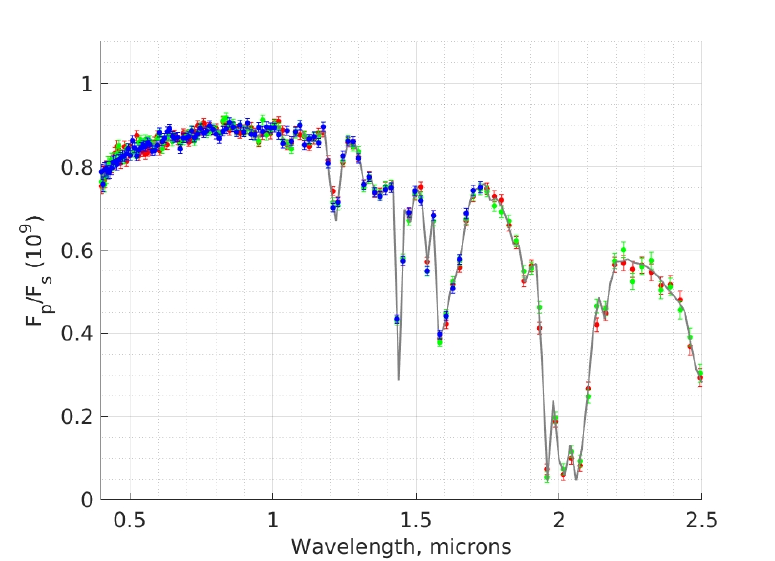}\label{fig:Venus_LUVOIR-b}}
\caption{Simulated spectral observation of a Venus analogue at 10 pc using a telescope with LUVOIR characteristics and an integration time of 10 hours. The parameters used for the telescope are shown in Table \ref{tab:model_LUVOIR}. The black line represents the noise-free spectrum from the 3D simulation, and corresponds to the planet phase orbit 0.25. The two panels correspond to different sizes of the telescope diameter: (a) is 8 meters and (b) is 15 meters. The different point colors are the results of the simulated observations with different inner working angles of the coronograph: $\lambda/$D (red), $2\lambda/$D (green) and $3\lambda/$D (blue).}
\label{fig:Venus_LUVOIR}
\end{centering}
\end{figure}

\begin{figure}
\begin{centering}
\subfigure[Telescope diameter, 8 meters]{
\includegraphics[width=1.0\columnwidth]{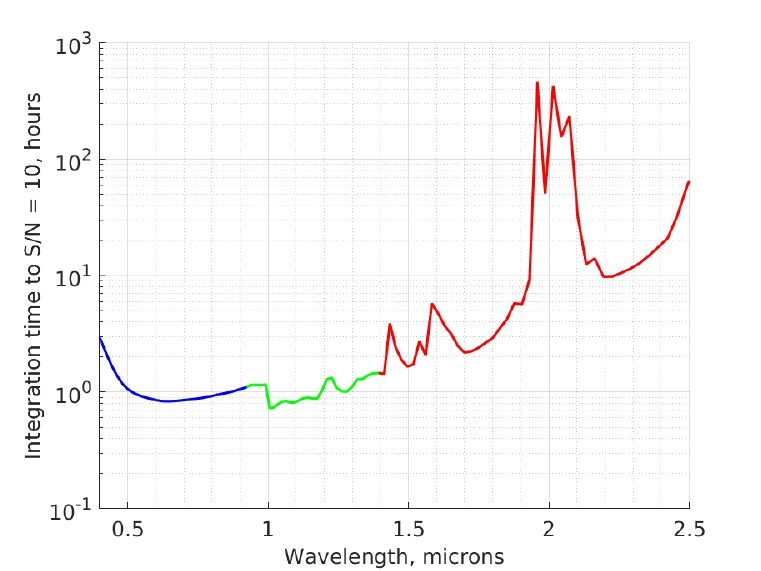}\label{fig:Venus_LUVOIR_Time-a}}
\subfigure[Telescope diameter, 15 meters]{
\includegraphics[width=1.0\columnwidth]{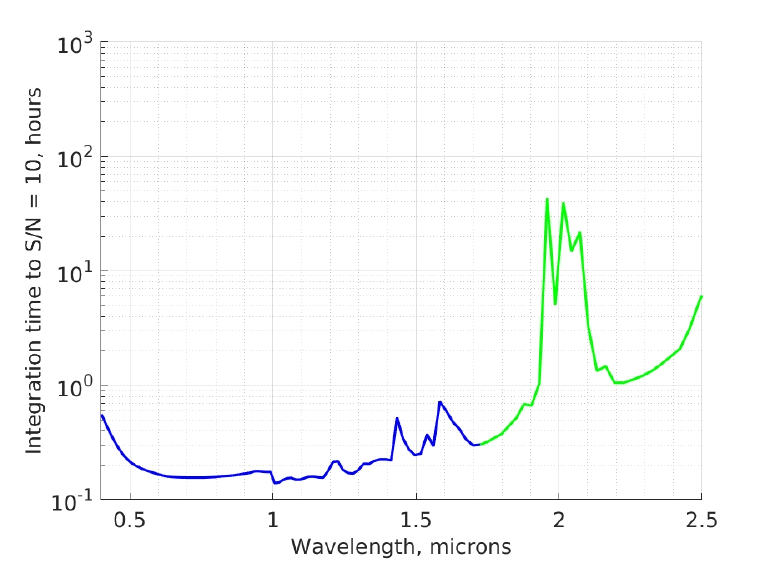}\label{fig:Venus_LUVOIR_Time-b}}
\caption{Integration time for a S/N equal to 10. This figure uses the same parameters as in Fig. \ref{fig:Venus_LUVOIR}. The different panels correspond to telescopes with different diameter ((a), 8 meters; (b), 15 meters). The different line colors are the results of the simulated observations with different inner working angles of the coronograph: $\lambda/$D (red), $2\lambda/$D (green) and $3\lambda/$D (blue). Note that in panel (b) the red color is not visible because it overlaps completely with the green line for this wavelength range (0.4-2.5 $\mu$m).}
\label{fig:Venus_LUVOIR_Time}
\end{centering}
\end{figure}

As shown in Fig. \ref{fig:spec_planet}, in order to be able to observe a Venus analogue, our telescope needs to be able to resolve changes in the planet flux at the parts-per-billion (ppb) level or smaller. Mission concepts such as LUVOIR \citep{2018LUVOIR} or HabEx  \citep{2018Habex} would make observations of Venus analogues feasible. Following the formalism developed in \cite{2016Robinson}, we have written a module in the \texttt{OASIS} platform that allow us to simulate an observational spectrum using directly the results from the 3D simulations. Our observed spectra are based on a telescope that incorporates a coronograph. The main parameters of the instrument are listed in Table \ref{tab:model_LUVOIR}. In Fig. \ref{fig:Venus_LUVOIR} we show the observed spectra at orbital phase 0.25 for different telescope diameters and 10 hours of observation time. The Venus analogue was placed 10 pc away as for the results shown in the previous figure (Fig. \ref{fig:spec_planet}). The spectral range observed is the same as the one proposed for the spectrograph in the LUVOIR concept mission. The instrument parameters are based on the work of \cite{2016Robinson}. It is known that small inner working angles for the coronograph in a large telescope such as LUVOIR is challenging due to telescope stability problems \citep{2016Robinson}. Additional information regarding the limitation of small angle coronographic techniques can be found in e.g., \cite{2012Mawet} and \cite{2014Mawet}. However, a large inner working angle make the observations of planets close to the star difficult. Note that the inner working angle is proportional to $\lambda/$D, where D is the telescope diameter. As in \cite{2016Robinson}, we have adopted the parameters for the near-infrared detector that represent the HgCdTe detectors (see Table \ref{tab:model_LUVOIR} and \citealt{2015Morgan}). In Fig. \ref{fig:Venus_LUVOIR-a}, we simulate the observations from a 8 meter LUVOIR telescope with different inner working angles. The blue points represent an inner working angle of $3\lambda/$D, which would only allow for observations at wavelengths shorter than 1 $\mu$m. It would be difficult to characterize the atmosphere from the data corresponding to the blue points, where the only information observed would be the reflected light from the cloud structure. With an inner working angle of $2\lambda/$D (green points) we would be able to capture the H$_2$O feature at roughly 1.2 $\mu$m. The red points represent an inner working angle of $\lambda/$D. By using this latter configuration in an 8 meter telescope we would be able to cover the instrument spectral range and capture the H$_2$O feature at 1.4$\mu$m plus the CO$_2$ features at 1.6 and 2 $\mu$m. The information gathered from the observed spectrum would be able to constrain the upper atmosphere and cloud region. To reconstruct the global atmosphere based on the information obtained from the upper atmosphere we would have to use theoretical models to make predictions on the deep atmospheric conditions. Fig. \ref{fig:Venus_LUVOIR-b} shows that a 15 meter diameter telescope would significantly improve the performance of the instrument, it would make it possible to observe the H$_2$O and CO$_2$ features with all the inner working angles tested. In Fig. \ref{fig:Venus_LUVOIR_Time} we show the roughly one order magnitude shorter time integration needed to acquire a S/N of 10 with a 15 meters diameter telescope compared with the 8 meters telescope. S/N is the signal-to-noise ratio defined by the ratio between the planet flux and the noise per spectral band. The maximum integration time needed is roughly 40 hours for the deep feature at 2 $\mu$m with the 15 meter telescope.

\begin{figure}
\begin{centering}
\subfigure[Telescope diameter, 8 meters]{
\includegraphics[width=1.0\columnwidth]{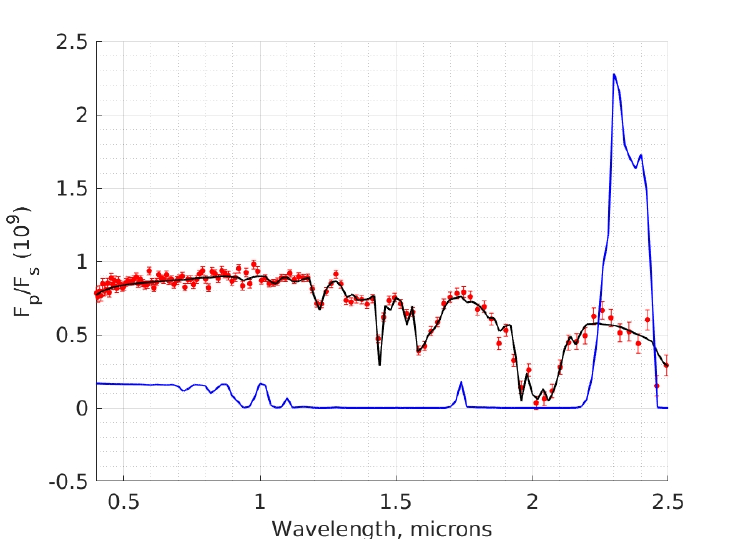}\label{fig:Venus_LUVOIR-a}}
\caption{Simulated spectral observation of a Venus analogue at 10 pc using a 8 m telescope with LUVOIR characteristics and an integration time of 10 hours. The parameters used for the telescope are shown in Table \ref{tab:model_LUVOIR}. The black line represents the noise-free spectrum from the reference simulation and blue line from the cloud free simulation, and both spectra correspond to the planet phase orbit 0.25. The inner working angles of the coronograph is defined as $3\lambda/$D.}
\label{fig:Venus_LUVOIR_cloudy_clear}
\end{centering}
\end{figure}

Fig. \ref{fig:Venus_LUVOIR_cloudy_clear} compares the spectrum of the reference simulation to the simulation with no clouds or unknown UV absorber, and shows that with the 8 meters LUVOIR (and 10 hours integration), it is possible to distinguish between a cloudy and a clear atmosphere. The highest peak in the clear atmosphere spectrum is due to the CO$_2$ emission from the hot deep layers, which in the cloudy atmosphere is absorbed at the cloud base.

\section{Conclusions and Future prospects}
\label{sec:conc}
The number of terrestrial planets discovered outside the Solar System with Earth-like masses and sizes continues to grow and there is a need of theoretical tools capable of characterizing the environment of these planets. We have developed a 3D platform, \texttt{OASIS}, capable of simulating the environment of terrestrial planets. The model has been developed from the ground up to avoid approximations that could hinder its flexibility to explore a large diversity of planetary conditions. The code has been developed using state-of-the-art programming techniques and graphic processing units (GPUs), currently among the best hardware to power high performance computing. In this work, we have coupled the dynamics code (\texttt{THOR}) with the new radiation code (\texttt{CYCLOPS}), a simplified representation of surface and soil thermodynamics and physical properties (\texttt{ATLANTIS}), and convection and cascade of enstrophy (\texttt{LOKI}). The cloud and chemistry parameters were kept constant during the simulations.

\texttt{OASIS} has completed its first benchmark test, where the modules were combined to simulate a terrestrial planet. We decided to model a planet with Venus-like conditions, both for its inherent complexity and for its relevance for future exoplanet discoveries. It is important to explore Venus-like planets in the context of exoplanet research since they will inform us about the conditions of the inner boundary of the habitable zone. They are also important from an observational point of view since the transit method will be biased towards the detection of planets with orbits closer to the host stars than farther away (\citealt{2008Kane}). The Venus environment is known to be technically and computationally difficult to simulate due to the complex optical structure of the atmosphere that requires a computationally expensive radiative transfer code coupled to a dynamical core to produce a robust 3D temperature structure and atmospheric circulation. Therefore Venus provides a challenging case to test the capabilities of the current \texttt{OASIS} platform.

Our GPU implementation allowed us to use the radiation scheme with a fairly good spectral resolution in a GCM (353 spectral band and 20 Gaussian points for the k-distribution optical data). The new code performed robust simulations of the massive Venus atmosphere and is able to successfully reproduce an atmospheric circulation and temperature structure similar to the observations of Venus.
 This is an important achievement and shows that our model can confidently be used to explore Venus-like planets in the future. The winds in the upper cloud region obtained in the simulations with 2 degree resolution are as strong as indicated by the observations from cloud tracking measurements. Our 2 degree resolution simulation was able to simulate these strong winds with no fine tuning. The strong winds are part of a phenomenon called the ``super-rotating'' Venus atmosphere. We found that when using lower space resolution models, the winds in the cloud region are weaker than when using 2 degrees resolution. The lower velocity wind values may be associated with the inaccuracies in the dynamical core when representing the angular momentum exchanges in the deep atmosphere for crude spatial resolutions (e.g., \citealt{2012Lebonnois}). Using higher spatial resolution helps reducing the numerical errors associated with the angular momentum in Venus simulations. With higher spatial resolution, such as 2 degrees, these numerical errors become substantially smaller compared to the physical sources and transport of angular momentum in the simulated atmosphere, as we show here. The model finds a warmer cloud region than in the observations, but this is expected and caused by the cloud distribution used. We used such cloud distribution to be consistent with previous work (e.g., \citealt{2011Lee}).  

We have also tested the impact of different model parameters on the atmospheric circulation and temperature. In summary, a) when the clouds are removed, stellar energy is deposited deeper in the atmosphere leading to stronger atmospheric cells and weaker zonal winds. Removing the clouds also has an important impact on the temperature structure, where the surface temperature increases and produces a deep convective region represented by an adiabatic profile extending from the surface to pressure levels as high as 100 mbar; b) Decreasing the heat capacity of the atmosphere to 700 J K$^{-1}$kg$^{-1}$ did not have a large impact on the simulated atmosphere. However, if the heat capacity is increased to 1100 J K$^{-1}$kg$^{-1}$ it has a large impact on the circulation produced, in the sense that the large scale cells become stronger, but the low latitude jet weakens significantly. Both experiments produced temperature structures similar to the reference simulation, but the pole-equator differences are smaller in the case of higher heat capacity. c) Reducing the surface friction by one order of magnitude did not have a significant impact on the simulated atmosphere, which was largely similar to the reference simulation.

One of the goals of \texttt{OASIS} is to guide us in the preparation of observations. We showed the results of simulated spectra of a Venus analogue at 10 pc away from Earth as observed with a telescope similar to the LUVOIR mission concept. To be able to observe this planet-star system at 10 pc distance we need to be able to reach values of planet/star contrasts of the order of 10$^{-10}$. The results show that it would be possible to characterize a Venus-like planet using LUVOIR, with a coronograph, a mirror of 8 meters and an inner working angle of $\lambda/$D. These results are consistent with the results presented in \cite{2018LUVOIR}. A specific configuration of the telescope diameter and the inner working angle of the coronograph is needed, and if the inner working angle is not sufficiently small, observations of Venus analogues will be almost impossible.

We have demonstrated that \texttt{OASIS} is capable of simulating terrestrial planet environments. As a platform, \texttt{OASIS} can be continuously updated by testing new algorithms and physical/chemical models to increase the sophistication and complexity of the platform. 

\section*{Acknowledgements}
We thank Sandra Raimundo and Tais Dahl for instructive conversations and comments on the manuscript. J.M. and L.A.B. acknowledge financial support from the VILLUM Foundation YIP Program and the Carlsberg Foundation Distinguished Associate Professor Fellowship. The results of this work were computed in the new GPU cluster (FOUNDATION) supported by the DTU Space's strategic funds 2018. 

%%%%%%%%%%%%%%%%%%%%%%%%%%%%%%%%%%%%%%%%%%%%%%%%%%

%%%%%%%%%%%%%%%%%%%% REFERENCES %%%%%%%%%%%%%%%%%%
\bibliographystyle{mnras}
%\bibliography{mybibfile} % if your bibtex file is called example.bib

\begin{thebibliography}{}
\makeatletter
\relax
\def\mn@urlcharsother{\let\do\@makeother \do\$\do\&\do\#\do\^\do\_\do\%\do\~}
\def\mn@doi{\begingroup\mn@urlcharsother \@ifnextchar [ {\mn@doi@}
  {\mn@doi@[]}}
\def\mn@doi@[#1]#2{\def\@tempa{#1}\ifx\@tempa\@empty \href
  {http://dx.doi.org/#2} {doi:#2}\else \href {http://dx.doi.org/#2} {#1}\fi
  \endgroup}
\def\mn@eprint#1#2{\mn@eprint@#1:#2::\@nil}
\def\mn@eprint@arXiv#1{\href {http://arxiv.org/abs/#1} {{\tt arXiv:#1}}}
\def\mn@eprint@dblp#1{\href {http://dblp.uni-trier.de/rec/bibtex/#1.xml}
  {dblp:#1}}
\def\mn@eprint@#1:#2:#3:#4\@nil{\def\@tempa {#1}\def\@tempb {#2}\def\@tempc
  {#3}\ifx \@tempc \@empty \let \@tempc \@tempb \let \@tempb \@tempa \fi \ifx
  \@tempb \@empty \def\@tempb {arXiv}\fi \@ifundefined
  {mn@eprint@\@tempb}{\@tempb:\@tempc}{\expandafter \expandafter \csname
  mn@eprint@\@tempb\endcsname \expandafter{\@tempc}}}

\bibitem[\protect\citeauthoryear{{Abbot} \& {Pierrehumbert}}{{Abbot} \&
  {Pierrehumbert}}{2010}]{2010Abbot}
{Abbot} D.~S.,  {Pierrehumbert} R.~T.,  2010, \mn@doi [Journal of Geophysical
  Research (Atmospheres)] {10.1029/2009JD012007}, \href
  {https://ui.adsabs.harvard.edu/abs/2010JGRD..115.3104A} {115, D03104}

\bibitem[\protect\citeauthoryear{{Allen} \& {Crawford}}{{Allen} \&
  {Crawford}}{1984}]{1984Allen}
{Allen} D.~A.,  {Crawford} J.~W.,  1984, \mn@doi [\nat] {10.1038/307222a0},
  \href {http://adsabs.harvard.edu/abs/1984Natur.307..222A} {307, 222}

\bibitem[\protect\citeauthoryear{{Anglada-Escud{\'e}}
  et~al.,}{{Anglada-Escud{\'e}} et~al.}{2016}]{2016Anglada}
{Anglada-Escud{\'e}} G.,  et~al., 2016, \mn@doi [\nat] {10.1038/nature19106},
  \href {https://ui.adsabs.harvard.edu/abs/2016Natur.536..437A} {536, 437}

\bibitem[\protect\citeauthoryear{{Baranov}, {Lafferty}  \& {Fraser}}{{Baranov}
  et~al.}{2004}]{2004Baranov}
{Baranov} Y.~I.,  {Lafferty} W.~J.,   {Fraser} G.~T.,  2004, \mn@doi [Journal
  of Molecular Spectroscopy] {10.1016/j.jms.2004.04.010}, \href
  {http://adsabs.harvard.edu/abs/2004JMoSp.228..432B} {228, 432}

\bibitem[\protect\citeauthoryear{{Batalha}, {Lewis}, {Line}, {Valenti}  \&
  {Stevenson}}{{Batalha} et~al.}{2018}]{2018Batalha}
{Batalha} N.~E.,  {Lewis} N.~K.,  {Line} M.~R.,  {Valenti} J.,   {Stevenson}
  K.,  2018, \mn@doi [\apjl] {10.3847/2041-8213/aab896}, \href
  {https://ui.adsabs.harvard.edu/abs/2018ApJ...856L..34B} {856, L34}

\bibitem[\protect\citeauthoryear{{B{\'e}zard}, {Fedorova}, {Bertaux}, {Rodin}
  \& {Korablev}}{{B{\'e}zard} et~al.}{2011}]{2011Bezard}
{B{\'e}zard} B.,  {Fedorova} A.,  {Bertaux} J.-L.,  {Rodin} A.,   {Korablev}
  O.,  2011, \mn@doi [\icarus] {10.1016/j.icarus.2011.08.025}, \href
  {http://adsabs.harvard.edu/abs/2011Icar..216..173B} {216, 173}

\bibitem[\protect\citeauthoryear{{Bonfils} et~al.,}{{Bonfils}
  et~al.}{2018}]{2018Bonfils}
{Bonfils} X.,  et~al., 2018, \mn@doi [\aap] {10.1051/0004-6361/201731973},
  \href {https://ui.adsabs.harvard.edu/abs/2018A%26A...613A..25B} {613, A25}

\bibitem[\protect\citeauthoryear{{Bullock} \& {Grinspoon}}{{Bullock} \&
  {Grinspoon}}{2001}]{2001Bullock}
{Bullock} M.~A.,  {Grinspoon} D.~H.,  2001, \mn@doi [\icarus]
  {10.1006/icar.2000.6570}, \href
  {https://ui.adsabs.harvard.edu/abs/2001Icar..150...19B} {150, 19}

\bibitem[\protect\citeauthoryear{{Caldas}, {Leconte}, {Selsis}, {Waldmann},
  {Bord{\'e}}, {Rocchetto}  \& {Charnay}}{{Caldas} et~al.}{2019}]{2019Caldas}
{Caldas} A.,  {Leconte} J.,  {Selsis} F.,  {Waldmann} I.~P.,  {Bord{\'e}} P.,
  {Rocchetto} M.,   {Charnay} B.,  2019, \mn@doi [\aap]
  {10.1051/0004-6361/201834384}, \href
  {https://ui.adsabs.harvard.edu/abs/2019A&A...623A.161C} {623, A161}

\bibitem[\protect\citeauthoryear{{Carpenter}}{{Carpenter}}{1964}]{1964Carpenter}
{Carpenter} R.~L.,  1964, \mn@doi [\aj] {10.1086/109220}, \href
  {http://adsabs.harvard.edu/abs/1964AJ.....69....2C} {69, p. 2}

\bibitem[\protect\citeauthoryear{{Catling} \& {Kasting}}{{Catling} \&
  {Kasting}}{2017}]{2017Catling}
{Catling} D.~C.,  {Kasting} J.~F.,  2017, {Atmospheric Evolution on Inhabited
  and Lifeless Worlds}

\bibitem[\protect\citeauthoryear{{Charnay}, {Forget}, {Wordsworth}, {Leconte},
  {Millour}, {Codron}  \& {Spiga}}{{Charnay} et~al.}{2013}]{2013Charnay}
{Charnay} B.,  {Forget} F.,  {Wordsworth} R.,  {Leconte} J.,  {Millour} E.,
  {Codron} F.,   {Spiga} A.,  2013, \mn@doi [Journal of Geophysical Research
  (Atmospheres)] {10.1002/jgrd.50808}, \href
  {https://ui.adsabs.harvard.edu/abs/2013JGRD..11810414C} {118, 10,414}

\bibitem[\protect\citeauthoryear{{Charnay}, {Barth}, {Rafkin}, {Narteau},
  {Lebonnois}, {Rodriguez}, {Courrech Du Pont}  \& {Lucas}}{{Charnay}
  et~al.}{2015}]{2015Charnay}
{Charnay} B.,  {Barth} E.,  {Rafkin} S.,  {Narteau} C.,  {Lebonnois} S.,
  {Rodriguez} S.,  {Courrech Du Pont} S.,   {Lucas} A.,  2015, \mn@doi [Nature
  Geoscience] {10.1038/ngeo2406}, \href
  {https://ui.adsabs.harvard.edu/abs/2015NatGe...8..362C} {8, 362}

\bibitem[\protect\citeauthoryear{{Crisp}}{{Crisp}}{1986}]{1986Crisp}
{Crisp} D.,  1986, \mn@doi [Icarus] {10.1016/0019-1035(86)90126-0}, \href
  {http://adsabs.harvard.edu/abs/1986Icar...67..484C} {67, 484}

\bibitem[\protect\citeauthoryear{{Crisp}}{{Crisp}}{1989}]{1989Crisp}
{Crisp} D.,  1989, \mn@doi [Icarus] {10.1016/0019-1035(89)90096-1}, \href
  {http://adsabs.harvard.edu/abs/1989Icar...77..391C} {77, 391}

\bibitem[\protect\citeauthoryear{{Deitrick}, {Mendon{\c{c}}a},
  {Schroffenegger}, {Grimm}, {Tsai}  \& {Heng}}{{Deitrick}
  et~al.}{2019}]{2019Deitrick}
{Deitrick} R.,  {Mendon{\c{c}}a} J.~M.,  {Schroffenegger} U.,  {Grimm} S.~L.,
  {Tsai} S.-M.,   {Heng} K.,  2019, arXiv e-prints, \href
  {https://ui.adsabs.harvard.edu/abs/2019arXiv191113158D} {p. arXiv:1911.13158}

\bibitem[\protect\citeauthoryear{{Del Genio}, {Way}, {Amundsen}, {Aleinov},
  {Kelley}, {Kiang}  \& {Clune}}{{Del Genio} et~al.}{2019}]{2019DelGenio}
{Del Genio} A.~D.,  {Way} M.~J.,  {Amundsen} D.~S.,  {Aleinov} I.,  {Kelley}
  M.,  {Kiang} N.~Y.,   {Clune} T.~L.,  2019, \mn@doi [Astrobiology]
  {10.1089/ast.2017.1760}, \href
  {https://ui.adsabs.harvard.edu/abs/2019AsBio..19...99D} {19, 99}

\bibitem[\protect\citeauthoryear{{Dittmann} et~al.,}{{Dittmann}
  et~al.}{2017}]{2017Dittmann}
{Dittmann} J.~A.,  et~al., 2017, \mn@doi [\nat] {10.1038/nature22055}, \href
  {https://ui.adsabs.harvard.edu/abs/2017Natur.544..333D} {544, 333}

\bibitem[\protect\citeauthoryear{Donner \& Large}{Donner \&
  Large}{2008}]{2008Donner}
Donner L.~J.,  Large W.~G.,  2008, \mn@doi [Annual Review of Environment and
  Resources] {10.1146/annurev.environ.33.020707.160752}, 33, 1

\bibitem[\protect\citeauthoryear{{Dressing} \& {Charbonneau}}{{Dressing} \&
  {Charbonneau}}{2015}]{2015Dressing}
{Dressing} C.~D.,  {Charbonneau} D.,  2015, \mn@doi [\apj]
  {10.1088/0004-637X/807/1/45}, \href
  {https://ui.adsabs.harvard.edu/abs/2015ApJ...807...45D} {807, 45}

\bibitem[\protect\citeauthoryear{{Eymet}, {Fournier}, {Dufresne}, {Lebonnois},
  {Hourdin}  \& {Bullock}}{{Eymet} et~al.}{2009}]{2009Eymet}
{Eymet} V.,  {Fournier} R.,  {Dufresne} J.,  {Lebonnois} S.,  {Hourdin} F.,
  {Bullock} M.~A.,  2009, \mn@doi [Journal of Geophysical Research (Planets)]
  {10.1029/2008JE003276}, \href
  {http://adsabs.harvard.edu/abs/2009JGRE..11411008E} {114, E11008}

\bibitem[\protect\citeauthoryear{{Fisher} \& {Heng}}{{Fisher} \&
  {Heng}}{2018}]{2018Fisher}
{Fisher} C.,  {Heng} K.,  2018, \mn@doi [\mnras] {10.1093/mnras/sty2550}, \href
  {http://adsabs.harvard.edu/abs/2018MNRAS.481.4698F} {481, 4698}

\bibitem[\protect\citeauthoryear{{Forget}}{{Forget}}{2013}]{2013Forget}
{Forget} F.,  2013, \mn@doi [International Journal of Astrobiology]
  {10.1017/S1473550413000128}, \href
  {https://ui.adsabs.harvard.edu/abs/2013IJAsB..12..177F} {12, 177}

\bibitem[\protect\citeauthoryear{{Forget} et~al.,}{{Forget}
  et~al.}{1999}]{1999Forget}
{Forget} F.,  et~al., 1999, \mn@doi [\jgr] {10.1029/1999JE001025}, \href
  {http://adsabs.harvard.edu/abs/1999JGR...10424155F} {104, 24155}

\bibitem[\protect\citeauthoryear{{Forget}, {Bertrand}, {Vangvichith},
  {Leconte}, {Millour}  \& {Lellouch}}{{Forget} et~al.}{2017}]{2017Forget}
{Forget} F.,  {Bertrand} T.,  {Vangvichith} M.,  {Leconte} J.,  {Millour} E.,
  {Lellouch} E.,  2017, \mn@doi [\icarus] {10.1016/j.icarus.2016.11.038}, \href
  {https://ui.adsabs.harvard.edu/abs/2017Icar..287...54F} {287, 54}

\bibitem[\protect\citeauthoryear{{Frandsen}, {Wennberg}  \&
  {Kjaergaard}}{{Frandsen} et~al.}{2016}]{2016Frandsen}
{Frandsen} B.~N.,  {Wennberg} P.~O.,   {Kjaergaard} H.~G.,  2016, \mn@doi
  [\grl] {10.1002/2016GL070916}, \href
  {http://adsabs.harvard.edu/abs/2016GeoRL..4311146F} {43, 11}

\bibitem[\protect\citeauthoryear{{Gamache} et~al.,}{{Gamache}
  et~al.}{2017}]{2017Gamache}
{Gamache} R.~R.,  et~al., 2017, \mn@doi [\jqsrt] {10.1016/j.jqsrt.2017.03.045},
  \href {http://adsabs.harvard.edu/abs/2017JQSRT.203...70G} {203, 70}

\bibitem[\protect\citeauthoryear{{Gaudi} et~al.,}{{Gaudi}
  et~al.}{2018}]{2018Habex}
{Gaudi} B.~S.,  et~al., 2018, arXiv e-prints -
  https://arxiv.org/abs/1809.09674, \href
  {https://ui.adsabs.harvard.edu/abs/2018arXiv180909674G} {}

\bibitem[\protect\citeauthoryear{{Gillon} et~al.,}{{Gillon}
  et~al.}{2016}]{2016Gillon}
{Gillon} M.,  et~al., 2016, \mn@doi [\nat] {10.1038/nature17448}, \href
  {https://ui.adsabs.harvard.edu/abs/2016Natur.533..221G} {533, 221}

\bibitem[\protect\citeauthoryear{{Gillon} et~al.,}{{Gillon}
  et~al.}{2017}]{2017Gillon}
{Gillon} M.,  et~al., 2017, \mn@doi [\nat] {10.1038/nature21360}, \href
  {https://ui.adsabs.harvard.edu/abs/2017Natur.542..456G} {542, 456}

\bibitem[\protect\citeauthoryear{{Goldstein}}{{Goldstein}}{1964}]{1964Goldstein}
{Goldstein} R.~M.,  1964, \mn@doi [\aj] {10.1086/109221}, \href
  {http://adsabs.harvard.edu/abs/1964AJ.....69...12G} {69, p. 12}

\bibitem[\protect\citeauthoryear{{Grimm} \& {Heng}}{{Grimm} \&
  {Heng}}{2015}]{2015Grimm}
{Grimm} S.~L.,  {Heng} K.,  2015, \mn@doi [\apj] {10.1088/0004-637X/808/2/182},
  \href {http://adsabs.harvard.edu/abs/2015ApJ...808..182G} {808, 182}

\bibitem[\protect\citeauthoryear{{Gruszka} \& {Borysow}}{{Gruszka} \&
  {Borysow}}{1997}]{1997Gruszka}
{Gruszka} M.,  {Borysow} A.,  1997, \mn@doi [\icarus] {10.1006/icar.1997.5773},
  \href {http://adsabs.harvard.edu/abs/1997Icar..129..172G} {129, 172}

\bibitem[\protect\citeauthoryear{{Haqq-Misra}, {Wolf}, {Joshi}, {Zhang}  \&
  {Kopparapu}}{{Haqq-Misra} et~al.}{2018}]{2018HaqqMisra}
{Haqq-Misra} J.,  {Wolf} E.~T.,  {Joshi} M.,  {Zhang} X.,   {Kopparapu} R.~K.,
  2018, \mn@doi [\apj] {10.3847/1538-4357/aa9f1f}, \href
  {https://ui.adsabs.harvard.edu/abs/2018ApJ...852...67H} {852, 67}

\bibitem[\protect\citeauthoryear{{Haus}, {Kappel}  \& {Arnold}}{{Haus}
  et~al.}{2015}]{2015Haus}
{Haus} R.,  {Kappel} D.,   {Arnold} G.,  2015, \mn@doi [\planss]
  {10.1016/j.pss.2015.06.024}, \href
  {http://adsabs.harvard.edu/abs/2015P%26SS..117..262H} {117, 262}

\bibitem[\protect\citeauthoryear{{Held} \& {Suarez}}{{Held} \&
  {Suarez}}{1994}]{1994Held}
{Held} I.~M.,  {Suarez} M.~J.,  1994, \mn@doi [Bulletin of the American
  Meteorological Society] {10.1175/1520-0477(1994)075<1825:APFTIO>2.0.CO;2},
  \href {http://adsabs.harvard.edu/abs/1994BAMS...75.1825H} {75, 1825}

\bibitem[\protect\citeauthoryear{{Hourdin}, {Le van}, {Forget}  \&
  {Talagrand}}{{Hourdin} et~al.}{1993}]{1993Hourdin}
{Hourdin} F.,  {Le van} P.,  {Forget} F.,   {Talagrand} O.,  1993, \mn@doi
  [Journal of Atmospheric Sciences]
  {10.1175/1520-0469(1993)050<3625:MVATAS>2.0.CO;2}, \href
  {http://adsabs.harvard.edu/abs/1993JAtS...50.3625H} {50, 3625}

\bibitem[\protect\citeauthoryear{{Ingersoll}}{{Ingersoll}}{1969}]{1969Ingersoll}
{Ingersoll} A.~P.,  1969, \mn@doi [Journal of Atmospheric Sciences]
  {10.1175/1520-0469(1969)026<1191:TRGAHO>2.0.CO;2}, \href
  {http://adsabs.harvard.edu/abs/1969JAtS...26.1191I} {26, 1191}

\bibitem[\protect\citeauthoryear{{Kane} \& {von Braun}}{{Kane} \& {von
  Braun}}{2008}]{2008Kane}
{Kane} S.~R.,  {von Braun} K.,  2008, \mn@doi [\apj] {10.1086/592381}, \href
  {https://ui.adsabs.harvard.edu/abs/2008ApJ...689..492K} {689, 492}

\bibitem[\protect\citeauthoryear{{Kane}, {Kopparapu}  \&
  {Domagal-Goldman}}{{Kane} et~al.}{2014}]{2014Kane}
{Kane} S.~R.,  {Kopparapu} R.~K.,   {Domagal-Goldman} S.~D.,  2014, \mn@doi
  [\apjl] {10.1088/2041-8205/794/1/L5}, \href
  {https://ui.adsabs.harvard.edu/abs/2014ApJ...794L...5K} {794, L5}

\bibitem[\protect\citeauthoryear{{Kerzhanovich} \& {Limaye}}{{Kerzhanovich} \&
  {Limaye}}{1985}]{1985Kerzhanovich}
{Kerzhanovich} V.~V.,  {Limaye} S.~S.,  1985, \mn@doi [Advances in Space
  Research] {10.1016/0273-1177(85)90198-X}, \href
  {http://adsabs.harvard.edu/abs/1985AdSpR...5...59K} {5, 59}

\bibitem[\protect\citeauthoryear{{Kliore}, {Moroz}  \& {Keating}}{{Kliore}
  et~al.}{1985}]{1985Kliore}
{Kliore} A.~J.,  {Moroz} V.~I.,   {Keating} G.~M.,  1985, Advances in Space
  Research, \href {http://adsabs.harvard.edu/abs/1985AdSpR...5k....K} {5}

\bibitem[\protect\citeauthoryear{{Knollenberg} \& {Hunten}}{{Knollenberg} \&
  {Hunten}}{1980}]{1980Knollenberg}
{Knollenberg} R.~G.,  {Hunten} D.~M.,  1980, \mn@doi [Journal of Geophysical
  Research] {10.1029/JA085iA13p08039}, \href
  {http://adsabs.harvard.edu/abs/1980JGR....85.8039K} {85, 8039}

\bibitem[\protect\citeauthoryear{{Knutson} et~al.,}{{Knutson}
  et~al.}{2009}]{2009Knutson}
{Knutson} H.~A.,  et~al., 2009, \mn@doi [\apj] {10.1088/0004-637X/690/1/822},
  \href {http://adsabs.harvard.edu/abs/2009ApJ...690..822K} {690, 822}

\bibitem[\protect\citeauthoryear{{Komacek} \& {Abbot}}{{Komacek} \&
  {Abbot}}{2019}]{2019Komacek}
{Komacek} T.~D.,  {Abbot} D.~S.,  2019, \mn@doi [\apj]
  {10.3847/1538-4357/aafb33}, \href
  {https://ui.adsabs.harvard.edu/abs/2019ApJ...871..245K} {871, 245}

\bibitem[\protect\citeauthoryear{{Kopparapu}, {Wolf}, {Haqq-Misra}, {Yang},
  {Kasting}, {Meadows}, {Terrien}  \& {Mahadevan}}{{Kopparapu}
  et~al.}{2016}]{2016Kopparapu}
{Kopparapu} R.~k.,  {Wolf} E.~T.,  {Haqq-Misra} J.,  {Yang} J.,  {Kasting}
  J.~F.,  {Meadows} V.,  {Terrien} R.,   {Mahadevan} S.,  2016, \mn@doi [\apj]
  {10.3847/0004-637X/819/1/84}, \href
  {https://ui.adsabs.harvard.edu/abs/2016ApJ...819...84K} {819, 84}

\bibitem[\protect\citeauthoryear{{Krasnopolsky}}{{Krasnopolsky}}{2017}]{2017Krasnopolsky}
{Krasnopolsky} V.~A.,  2017, \mn@doi [\icarus] {10.1016/j.icarus.2016.10.003},
  \href {http://adsabs.harvard.edu/abs/2017Icar..286..134K} {286, 134}

\bibitem[\protect\citeauthoryear{{Lacis} \& {Oinas}}{{Lacis} \&
  {Oinas}}{1991}]{1991Lacis}
{Lacis} A.~A.,  {Oinas} V.,  1991, \mn@doi [\jgr] {10.1029/90JD01945}, \href
  {http://adsabs.harvard.edu/abs/1991JGR....96.9027L} {96, 9027}

\bibitem[\protect\citeauthoryear{{Lavie} et~al.,}{{Lavie}
  et~al.}{2017}]{2017Lavie}
{Lavie} B.,  et~al., 2017, \mn@doi [\aj] {10.3847/1538-3881/aa7ed8}, \href
  {https://ui.adsabs.harvard.edu/abs/2017AJ....154...91L} {154, 91}

\bibitem[\protect\citeauthoryear{{Lebonnois}, {Hourdin}, {Eymet}, {Crespin},
  {Fournier}  \& {Forget}}{{Lebonnois} et~al.}{2010}]{2010Lebonnois}
{Lebonnois} S.,  {Hourdin} F.,  {Eymet} V.,  {Crespin} A.,  {Fournier} R.,
  {Forget} F.,  2010, \mn@doi [Journal of Geophysical Research (Planets)]
  {10.1029/2009JE003458}, \href
  {http://adsabs.harvard.edu/abs/2010JGRE..11506006L} {115}

\bibitem[\protect\citeauthoryear{{Lebonnois}, {Covey}, {Grossman}, {Parish},
  {Schubert}, {Walterscheid}, {Lauritzen}  \& {Jablonowski}}{{Lebonnois}
  et~al.}{2012}]{2012Lebonnois}
{Lebonnois} S.,  {Covey} C.,  {Grossman} A.,  {Parish} H.,  {Schubert} G.,
  {Walterscheid} R.,  {Lauritzen} P.,   {Jablonowski} C.,  2012, \mn@doi
  [Journal of Geophysical Research (Planets)] {10.1029/2012JE004223}, \href
  {https://ui.adsabs.harvard.edu/abs/2012JGRE..11712004L} {117, E12004}

\bibitem[\protect\citeauthoryear{{Lebonnois}, {Eymet}, {Lee}  \& {Vatant
  d'Ollone}}{{Lebonnois} et~al.}{2015}]{2015Lebonnois}
{Lebonnois} S.,  {Eymet} V.,  {Lee} C.,   {Vatant d'Ollone} J.,  2015, \mn@doi
  [Journal of Geophysical Research (Planets)] {10.1002/2015JE004794}, \href
  {http://adsabs.harvard.edu/abs/2015JGRE..120.1186L} {120, 1186}

\bibitem[\protect\citeauthoryear{{Lebonnois}, {Sugimoto}  \&
  {Gilli}}{{Lebonnois} et~al.}{2016}]{2016Lebonnois}
{Lebonnois} S.,  {Sugimoto} N.,   {Gilli} G.,  2016, \mn@doi [\icarus]
  {10.1016/j.icarus.2016.06.004}, \href
  {https://ui.adsabs.harvard.edu/abs/2016Icar..278...38L} {278, 38}

\bibitem[\protect\citeauthoryear{{Lebrun}, {Massol}, {Chassefi{\`e}Re},
  {Davaille}, {Marcq}, {Sarda}, {Leblanc}  \& {Brandeis}}{{Lebrun}
  et~al.}{2013}]{2013Lebrun}
{Lebrun} T.,  {Massol} H.,  {Chassefi{\`e}Re} E.,  {Davaille} A.,  {Marcq} E.,
  {Sarda} P.,  {Leblanc} F.,   {Brandeis} G.,  2013, \mn@doi [Journal of
  Geophysical Research (Planets)] {10.1002/jgre.20068}, \href
  {http://adsabs.harvard.edu/abs/2013JGRE..118.1155L} {118, 1155}

\bibitem[\protect\citeauthoryear{{Leconte}, {Forget}, {Charnay}, {Wordsworth}
  \& {Pottier}}{{Leconte} et~al.}{2013}]{2013Leconte}
{Leconte} J.,  {Forget} F.,  {Charnay} B.,  {Wordsworth} R.,   {Pottier} A.,
  2013, \mn@doi [\nat] {10.1038/nature12827}, \href
  {https://ui.adsabs.harvard.edu/abs/2013Natur.504..268L} {504, 268}

\bibitem[\protect\citeauthoryear{{Lee} \& {Richardson}}{{Lee} \&
  {Richardson}}{2011}]{2011Lee}
{Lee} C.,  {Richardson} M.~I.,  2011, \mn@doi [Journal of Atmospheric Sciences]
  {10.1175/2011JAS3703.1}, \href
  {http://adsabs.harvard.edu/abs/2011JAtS...68.1323L} {68, 1323}

\bibitem[\protect\citeauthoryear{{Lee}, {Lewis}  \& {Read}}{{Lee}
  et~al.}{2007}]{2007Lee3}
{Lee} C.,  {Lewis} S.~R.,   {Read} P.~L.,  2007, \jgr, \href
  {http://adsabs.harvard.edu/abs/2007JGRE..11204S11L} {112 (E4), E04S11}

\bibitem[\protect\citeauthoryear{{Leovy}}{{Leovy}}{1973}]{1973Leovy}
{Leovy} C.~B.,  1973, Journal of Atmospheric Sciences, \href
  {http://adsabs.harvard.edu/abs/1973JAtS...30.1218L} {30, 1218}

\bibitem[\protect\citeauthoryear{{Limaye}, {Mogul}, {Smith}, {Ansari},
  {S{\l}owik}  \& {Vaishampayan}}{{Limaye} et~al.}{2018}]{2018Limaye}
{Limaye} S.~S.,  {Mogul} R.,  {Smith} D.~J.,  {Ansari} A.~H.,  {S{\l}owik}
  G.~P.,   {Vaishampayan} P.,  2018, \mn@doi [Astrobiology]
  {10.1089/ast.2017.1783}, \href
  {http://adsabs.harvard.edu/abs/2018AsBio..18.1181L} {18, 1181}

\bibitem[\protect\citeauthoryear{{Lincowski}, {Meadows}, {Crisp}, {Robinson},
  {Luger}, {Lustig-Yaeger}  \& {Arney}}{{Lincowski}
  et~al.}{2018}]{2018Lincowski}
{Lincowski} A.~P.,  {Meadows} V.~S.,  {Crisp} D.,  {Robinson} T.~D.,  {Luger}
  R.,  {Lustig-Yaeger} J.,   {Arney} G.~N.,  2018, \mn@doi [\apj]
  {10.3847/1538-4357/aae36a}, \href
  {https://ui.adsabs.harvard.edu/abs/2018ApJ...867...76L} {867, 76}

\bibitem[\protect\citeauthoryear{{Luger} et~al.,}{{Luger}
  et~al.}{2017}]{2017Luger}
{Luger} R.,  et~al., 2017, \mn@doi [Nature Astronomy]
  {10.1038/s41550-017-0129}, \href
  {https://ui.adsabs.harvard.edu/abs/2017NatAs...1E.129L} {1, 0129}

\bibitem[\protect\citeauthoryear{{Lustig-Yaeger}, {Meadows}  \&
  {Lincowski}}{{Lustig-Yaeger} et~al.}{2019}]{2019Lustig-Yaeger}
{Lustig-Yaeger} J.,  {Meadows} V.~S.,   {Lincowski} A.~P.,  2019, \mn@doi [\aj]
  {10.3847/1538-3881/ab21e0}, \href
  {https://ui.adsabs.harvard.edu/abs/2019AJ....158...27L} {158, 27}

\bibitem[\protect\citeauthoryear{{Malik} et~al.,}{{Malik}
  et~al.}{2017}]{2017Malik}
{Malik} M.,  et~al., 2017, \mn@doi [\aj] {10.3847/1538-3881/153/2/56}, \href
  {http://adsabs.harvard.edu/abs/2017AJ....153...56M} {153, 56}

\bibitem[\protect\citeauthoryear{{Malik}, {Kitzmann}, {Mendon{\c{c}}a},
  {Grimm}, {Marleau}, {Linder}, {Tsai}  \& {Heng}}{{Malik}
  et~al.}{2019}]{2019Malik}
{Malik} M.,  {Kitzmann} D.,  {Mendon{\c{c}}a} J.~M.,  {Grimm} S.~L.,  {Marleau}
  G.-D.,  {Linder} E.~F.,  {Tsai} S.-M.,   {Heng} K.,  2019, \mn@doi [\aj]
  {10.3847/1538-3881/ab1084}, \href
  {https://ui.adsabs.harvard.edu/abs/2019AJ....157..170M} {157, 170}

\bibitem[\protect\citeauthoryear{{Marcq}, {Encrenaz}, {B{\'e}zard}  \&
  {Birlan}}{{Marcq} et~al.}{2006}]{2006Marcq}
{Marcq} E.,  {Encrenaz} T.,  {B{\'e}zard} B.,   {Birlan} M.,  2006, \mn@doi
  [\planss] {10.1016/j.pss.2006.04.024}, \href
  {http://adsabs.harvard.edu/abs/2006P%26SS...54.1360M} {54, 1360}

\bibitem[\protect\citeauthoryear{{Mawet} et~al.,}{{Mawet}
  et~al.}{2012}]{2012Mawet}
{Mawet} D.,  et~al., 2012, in Space Telescopes and Instrumentation 2012:
  Optical, Infrared, and Millimeter Wave. p. 844204 (\mn@eprint {arXiv}
  {1207.5481}), \mn@doi{10.1117/12.927245}

\bibitem[\protect\citeauthoryear{{Mawet} et~al.,}{{Mawet}
  et~al.}{2014}]{2014Mawet}
{Mawet} D.,  et~al., 2014, \mn@doi [\apj] {10.1088/0004-637X/792/2/97}, \href
  {https://ui.adsabs.harvard.edu/abs/2014ApJ...792...97M} {792, 97}

\bibitem[\protect\citeauthoryear{{Mayne} et~al.,}{{Mayne}
  et~al.}{2017}]{2017Mayne}
{Mayne} N.~J.,  et~al., 2017, \mn@doi [\aap] {10.1051/0004-6361/201730465},
  \href {https://ui.adsabs.harvard.edu/abs/2017A%26A...604A..79M} {604, A79}

\bibitem[\protect\citeauthoryear{{Meadows} \& {Crisp}}{{Meadows} \&
  {Crisp}}{1996}]{1996Meadows}
{Meadows} V.~S.,  {Crisp} D.,  1996, \mn@doi [\jgr] {10.1029/95JE03567}, \href
  {http://adsabs.harvard.edu/abs/1996JGR...101.4595M} {101, 4595}

\bibitem[\protect\citeauthoryear{{Mendon{\c c}a} \& {Read}}{{Mendon{\c c}a} \&
  {Read}}{2016}]{2016Mendoncaa}
{Mendon{\c c}a} J.~M.,  {Read} P.~L.,  2016, \mn@doi [\planss]
  {10.1016/j.pss.2016.09.001}, \href
  {http://adsabs.harvard.edu/abs/2016P%26SS..134....1M} {134, 1}

\bibitem[\protect\citeauthoryear{{Mendon{\c c}a}, {Read}, {Wilson}  \&
  {Lewis}}{{Mendon{\c c}a} et~al.}{2012}]{2012Mendonca}
{Mendon{\c c}a} J.~M.,  {Read} P.~L.,  {Wilson} C.~F.,   {Lewis} S.~R.,  2012,
  \mn@doi [Icarus] {10.1016/j.icarus.2011.07.010}, \href
  {http://adsabs.harvard.edu/abs/2012Icar..217..629M} {217, 629}

\bibitem[\protect\citeauthoryear{{Mendon{\c c}a}, {Read}, {Wilson}  \&
  {Lee}}{{Mendon{\c c}a} et~al.}{2015}]{2015Mendonca}
{Mendon{\c c}a} J.~M.,  {Read} P.~L.,  {Wilson} C.~F.,   {Lee} C.,  2015,
  \mn@doi [\planss] {10.1016/j.pss.2014.11.008}, \href
  {http://adsabs.harvard.edu/abs/2015P%26SS..105...80M} {105, 80}

\bibitem[\protect\citeauthoryear{{Mendon{\c c}a}, {Grimm}, {Grosheintz}  \&
  {Heng}}{{Mendon{\c c}a} et~al.}{2016}]{2016Mendoncab}
{Mendon{\c c}a} J.~M.,  {Grimm} S.~L.,  {Grosheintz} L.,   {Heng} K.,  2016,
  \mn@doi [\apj] {10.3847/0004-637X/829/2/115}, \href
  {http://adsabs.harvard.edu/abs/2016ApJ...829..115M} {829, 115}

\bibitem[\protect\citeauthoryear{{Mendon{\c c}a}, {Tsai}, {Malik}, {Grimm}  \&
  {Heng}}{{Mendon{\c c}a} et~al.}{2018}]{2018Mendoncab}
{Mendon{\c c}a} J.~M.,  {Tsai} S.-m.,  {Malik} M.,  {Grimm} S.~L.,   {Heng} K.,
   2018, \mn@doi [\apj] {10.3847/1538-4357/aaed23}, \href
  {http://adsabs.harvard.edu/abs/2018ApJ...869..107M} {869, 107}

\bibitem[\protect\citeauthoryear{{Morgan} \& {Siegler}}{{Morgan} \&
  {Siegler}}{2015}]{2015Morgan}
{Morgan} R.,  {Siegler} N.,  2015, in Techniques and Instrumentation for
  Detection of Exoplanets VII. p. 96052I, \mn@doi{10.1117/12.2196770}

\bibitem[\protect\citeauthoryear{{Morley}, {Kreidberg}, {Rustamkulov},
  {Robinson}  \& {Fortney}}{{Morley} et~al.}{2017}]{2017Morley}
{Morley} C.~V.,  {Kreidberg} L.,  {Rustamkulov} Z.,  {Robinson} T.,   {Fortney}
  J.~J.,  2017, \mn@doi [\apj] {10.3847/1538-4357/aa927b}, \href
  {https://ui.adsabs.harvard.edu/abs/2017ApJ...850..121M} {850, 121}

\bibitem[\protect\citeauthoryear{{Palmer} \& {Williams}}{{Palmer} \&
  {Williams}}{1975}]{1975Palmer}
{Palmer} K.~F.,  {Williams} D.,  1975, \mn@doi [\ao] {10.1364/AO.14.000208},
  \href {http://adsabs.harvard.edu/abs/1975ApOpt..14..208P} {14, 208}

\bibitem[\protect\citeauthoryear{{Pieters}, {Head}, {Patterson}, {Pratt}  \&
  {Garvin}}{{Pieters} et~al.}{1986}]{1986Pieters}
{Pieters} C.~M.,  {Head} J.~W.,  {Patterson} W.,  {Pratt} S.,   {Garvin} J.,
  1986, \mn@doi [Science] {10.1126/science.234.4782.1379}, \href
  {http://adsabs.harvard.edu/abs/1986Sci...234.1379P} {234, 1379}

\bibitem[\protect\citeauthoryear{{Read}}{{Read}}{1986}]{1986Read}
{Read} P.~L.,  1986, \mn@doi [Quarterly Journal of the Royal Meteorological
  Society] {10.1256/smsqj.47113}, \href
  {http://adsabs.harvard.edu/abs/1986QJRMS.112..253R} {112, 253}

\bibitem[\protect\citeauthoryear{{Read} et~al.,}{{Read}
  et~al.}{2016}]{2016Read}
{Read} P.~L.,  et~al., 2016, \mn@doi [Quarterly Journal of the Royal
  Meteorological Society] {10.1002/qj.2704}, \href
  {https://ui.adsabs.harvard.edu/abs/2016QJRMS.142..703R} {142, 703}

\bibitem[\protect\citeauthoryear{{Read} et~al.,}{{Read}
  et~al.}{2017}]{2017Read}
{Read} P.~L.,  et~al., 2017, {The Martian Planetary Boundary Layer}.
pp 106--171, \mn@doi{10.1017/9781139060172.007}

\bibitem[\protect\citeauthoryear{{Rees}}{{Rees}}{1999}]{1999Rees}
{Rees} G.,  1999, Cambridge University Press, ISBN 052148040X, \href
  {http://adsabs.harvard.edu/abs/1999rsdb.book.....R} {}

\bibitem[\protect\citeauthoryear{{Robinson} \& {Catling}}{{Robinson} \&
  {Catling}}{2012}]{2012Robinson}
{Robinson} T.~D.,  {Catling} D.~C.,  2012, \mn@doi [\apj]
  {10.1088/0004-637X/757/1/104}, \href
  {https://ui.adsabs.harvard.edu/abs/2012ApJ...757..104R} {757, 104}

\bibitem[\protect\citeauthoryear{{Robinson}, {Stapelfeldt}  \&
  {Marley}}{{Robinson} et~al.}{2016}]{2016Robinson}
{Robinson} T.~D.,  {Stapelfeldt} K.~R.,   {Marley} M.~S.,  2016, \mn@doi
  [\pasp] {10.1088/1538-3873/128/960/025003}, \href
  {https://ui.adsabs.harvard.edu/abs/2016PASP..128b5003R} {128, 025003}

\bibitem[\protect\citeauthoryear{{Sagan}}{{Sagan}}{1961}]{1961Sagan}
{Sagan} C.,  1961, \mn@doi [Science] {10.1126/science.133.3456.849}, \href
  {http://adsabs.harvard.edu/abs/1961Sci...133..849S} {133, 849}

\bibitem[\protect\citeauthoryear{{S{\'a}nchez-Lavega}, {Lebonnois}, {Imamura},
  {Read}  \& {Luz}}{{S{\'a}nchez-Lavega} et~al.}{2017}]{2017Lavega}
{S{\'a}nchez-Lavega} A.,  {Lebonnois} S.,  {Imamura} T.,  {Read} P.,   {Luz}
  D.,  2017, \mn@doi [\ssr] {10.1007/s11214-017-0389-x}, \href
  {https://ui.adsabs.harvard.edu/abs/2017SSRv..212.1541S} {212, 1541}

\bibitem[\protect\citeauthoryear{{Satoh}}{{Satoh}}{2002}]{2002Satoh}
{Satoh} M.,  2002, \mn@doi [Monthly Weather Review]
  {10.1175/1520-0493(2002)130<1227:CSFTCN>2.0.CO;2}, \href
  {http://adsabs.harvard.edu/abs/2002MWRv..130.1227S} {130, 1227}

\bibitem[\protect\citeauthoryear{{Shields}}{{Shields}}{2019}]{2019Shields}
{Shields} A.~L.,  2019, \mn@doi [\apjs] {10.3847/1538-4365/ab2fe7}, \href
  {https://ui.adsabs.harvard.edu/abs/2019ApJS..243...30S} {243, 30}

\bibitem[\protect\citeauthoryear{{Shields}, {Barnes}, {Agol}, {Charnay}, {Bitz}
   \& {Meadows}}{{Shields} et~al.}{2016}]{2016Shields}
{Shields} A.~L.,  {Barnes} R.,  {Agol} E.,  {Charnay} B.,  {Bitz} C.,
  {Meadows} V.~S.,  2016, \mn@doi [Astrobiology] {10.1089/ast.2015.1353}, \href
  {https://ui.adsabs.harvard.edu/abs/2016AsBio..16..443S} {16, 443}

\bibitem[\protect\citeauthoryear{{Stamnes}, {Tsay}, {Wiscombe}  \&
  {Laszlo}}{{Stamnes} et~al.}{2000}]{2000Stamnes}
{Stamnes} K.,  {Tsay} S.,  {Wiscombe} W.,   {Laszlo} I.,  2000, Rep. available
  from ftp://climatel.gsfc.nasa.gov.wiscombe

\bibitem[\protect\citeauthoryear{{Taylor}, {Svedhem}  \& {Head}}{{Taylor}
  et~al.}{2018}]{2018Taylor}
{Taylor} F.~W.,  {Svedhem} H.,   {Head} J.~W.,  2018, \mn@doi [\ssr]
  {10.1007/s11214-018-0467-8}, \href
  {http://adsabs.harvard.edu/abs/2018SSRv..214...35T} {214, 35}

\bibitem[\protect\citeauthoryear{{Tellmann}, {P{\"a}tzold}, {H{\"a}usler},
  {Bird}  \& {Tyler}}{{Tellmann} et~al.}{2009}]{2009Tellmann}
{Tellmann} S.,  {P{\"a}tzold} M.,  {H{\"a}usler} B.,  {Bird} M.~K.,   {Tyler}
  G.~L.,  2009, \mn@doi [Journal of Geophysical Research (Planets)]
  {10.1029/2008JE003204}, \href
  {http://adsabs.harvard.edu/abs/2009JGRE..11400B36T} {114, p. E00B36}

\bibitem[\protect\citeauthoryear{{The LUVOIR Team}}{{The LUVOIR
  Team}}{2018}]{2018LUVOIR}
{The LUVOIR Team} 2018, arXiv e-prints - https://arxiv.org/pdf/1809.09668.pdf,
  \href {https://ui.adsabs.harvard.edu/abs/2018arXiv180909668T} {}

\bibitem[\protect\citeauthoryear{{Thuburn}}{{Thuburn}}{2008}]{2008Thuburn}
{Thuburn} J.,  2008, \mn@doi [Journal of Computational Physics]
  {10.1016/j.jcp.2006.08.016}, \href
  {http://adsabs.harvard.edu/abs/2008JCoPh.227.3715T} {227, 3715}

\bibitem[\protect\citeauthoryear{{Tomasko}, {Doose}, {Smith}  \&
  {Odell}}{{Tomasko} et~al.}{1980}]{1980Tomasko}
{Tomasko} M.~G.,  {Doose} L.~R.,  {Smith} P.~H.,   {Odell} A.~P.,  1980,
  \mn@doi [Journal of Geophysical Research] {10.1029/JA085iA13p08167}, \href
  {http://adsabs.harvard.edu/abs/1980JGR....85.8167T} {85, 8167}

\bibitem[\protect\citeauthoryear{{Tsai}, {Lyons}, {Grosheintz}, {Rimmer},
  {Kitzmann}  \& {Heng}}{{Tsai} et~al.}{2017}]{2017Tsai}
{Tsai} S.-M.,  {Lyons} J.~R.,  {Grosheintz} L.,  {Rimmer} P.~B.,  {Kitzmann}
  D.,   {Heng} K.,  2017, \mn@doi [\apjs] {10.3847/1538-4365/228/2/20}, \href
  {http://adsabs.harvard.edu/abs/2017ApJS..228...20T} {228, 20}

\bibitem[\protect\citeauthoryear{{Turbet}, {Forget}, {Leconte}, {Charnay}  \&
  {Tobie}}{{Turbet} et~al.}{2017}]{2017Turbet}
{Turbet} M.,  {Forget} F.,  {Leconte} J.,  {Charnay} B.,   {Tobie} G.,  2017,
  \mn@doi [Earth and Planetary Science Letters] {10.1016/j.epsl.2017.07.050},
  \href {https://ui.adsabs.harvard.edu/abs/2017E%26PSL.476...11T} {476, 11}

\bibitem[\protect\citeauthoryear{{Warrilow}, {Sangster}  \&
  {Slingo}}{{Warrilow} et~al.}{1986}]{1986Warrilow}
{Warrilow} D.~A.,  {Sangster} A.~B.,   {Slingo} A.,  1986, Tech. Rep. DCTN 38,
  UKMO, Meteorological Office, London Road, Bracknell, Bershire, United Kingdom

\bibitem[\protect\citeauthoryear{{Wolf} \& {Toon}}{{Wolf} \&
  {Toon}}{2013}]{2013Wolf}
{Wolf} E.~T.,  {Toon} O.~B.,  2013, \mn@doi [Astrobiology]
  {10.1089/ast.2012.0936}, \href
  {https://ui.adsabs.harvard.edu/abs/2013AsBio..13..656W} {13, 656}

\bibitem[\protect\citeauthoryear{Wordsworth}{Wordsworth}{2015}]{2015Wordsworth}
Wordsworth R.,  2015, \mn@doi [The Astrophysical Journal]
  {10.1088/0004-637x/806/2/180}, 806, 180

\bibitem[\protect\citeauthoryear{{Wordsworth}, {Forget}, {Selsis}, {Millour},
  {Charnay}  \& {Madeleine}}{{Wordsworth} et~al.}{2011}]{2011Wordsworth}
{Wordsworth} R.~D.,  {Forget} F.,  {Selsis} F.,  {Millour} E.,  {Charnay} B.,
  {Madeleine} J.-B.,  2011, \mn@doi [\apjl] {10.1088/2041-8205/733/2/L48},
  \href {http://adsabs.harvard.edu/abs/2011ApJ...733L..48W} {733, L48}

\bibitem[\protect\citeauthoryear{{Wordsworth}, {Kalugina}, {Lokshtanov},
  {Vigasin}, {Ehlmann}, {Head}, {Sanders}  \& {Wang}}{{Wordsworth}
  et~al.}{2017}]{2017Wordsworth}
{Wordsworth} R.,  {Kalugina} Y.,  {Lokshtanov} S.,  {Vigasin} A.,  {Ehlmann}
  B.,  {Head} J.,  {Sanders} C.,   {Wang} H.,  2017, \mn@doi [\grl]
  {10.1002/2016GL071766}, \href
  {https://ui.adsabs.harvard.edu/abs/2017GeoRL..44..665W} {44, 665}

\bibitem[\protect\citeauthoryear{{Wunderlich} et~al.,}{{Wunderlich}
  et~al.}{2019}]{2019Wunderlich}
{Wunderlich} F.,  et~al., 2019, \mn@doi [\aap] {10.1051/0004-6361/201834504},
  \href {https://ui.adsabs.harvard.edu/abs/2019A&A...624A..49W} {624, A49}

\bibitem[\protect\citeauthoryear{Yang, Abbot, Koll, Hu  \& Showman}{Yang
  et~al.}{2019}]{2019Yang}
Yang J.,  Abbot D.~S.,  Koll D. D.~B.,  Hu Y.,   Showman A.~P.,  2019, \mn@doi
  [The Astrophysical Journal] {10.3847/1538-4357/aaf1a8}, 871, 29

\bibitem[\protect\citeauthoryear{{Zasova}, {Moroz}, {Formisano}, {Ignatiev}  \&
  {Khatuntsev}}{{Zasova} et~al.}{2004}]{2004Zasova}
{Zasova} L.~V.,  {Moroz} V.~I.,  {Formisano} V.,  {Ignatiev} N.~I.,
  {Khatuntsev} I.~V.,  2004, \mn@doi [Advances in Space Research]
  {10.1016/j.asr.2003.09.067}, \href
  {https://ui.adsabs.harvard.edu/abs/2004AdSpR..34.1655Z} {34, 1655}

\bibitem[\protect\citeauthoryear{{Zhang} \& {Showman}}{{Zhang} \&
  {Showman}}{2017}]{2017Zhang}
{Zhang} X.,  {Showman} A.~P.,  2017, \mn@doi [\apj]
  {10.3847/1538-4357/836/1/73}, \href
  {https://ui.adsabs.harvard.edu/abs/2017ApJ...836...73Z} {836, 73}

\bibitem[\protect\citeauthoryear{{de Bergh}, {Bezard}, {Crisp}, {Maillard},
  {Owen}, {Pollack}  \& {Grinspoon}}{{de Bergh} et~al.}{1995}]{1995Bergh}
{de Bergh} C.,  {Bezard} B.,  {Crisp} D.,  {Maillard} J.~P.,  {Owen} T.,
  {Pollack} J.,   {Grinspoon} D.,  1995, \mn@doi [Advances in Space Research]
  {10.1016/0273-1177(94)00067-B}, \href
  {http://adsabs.harvard.edu/abs/1995AdSpR..15...79D} {15, 79}

\bibitem[\protect\citeauthoryear{{von Zahn} \& {Moroz}}{{von Zahn} \&
  {Moroz}}{1985}]{1985Zahn}
{von Zahn} U.,  {Moroz} V.~I.,  1985, \mn@doi [Advances in Space Research]
  {10.1016/0273-1177(85)90201-7}, \href
  {http://adsabs.harvard.edu/abs/1985AdSpR...5..173V} {5, 173}

\makeatother
\end{thebibliography}

%%%%%%%%%%%%%%%%%%%%%%%%%%%%%%%%%%%%%%%%%%%%%%%%%%

% Don't change these lines
\bsp	% typesetting comment
\label{lastpage}
\end{document}